\documentclass[structabstract]{aa} 
\usepackage{amsmath}
\usepackage{amstext}
\usepackage{amsfonts}
\usepackage{amssymb}
\usepackage{graphicx}
\usepackage[varg]{txfonts}
\usepackage{xspace}
\usepackage{aalongtable}

\usepackage{natbib}
\usepackage{multirow}

\usepackage{siunitx}
\begin{document} 

	\title{Large scale kinematics and dynamical modelling of the Milky Way nuclear star cluster\thanks{Based on observations collected at the European Organisation for Astronomical Research in the Southern 	Hemisphere, Chile \mbox{(289.B-5010(A)}).}$^{,}$\thanks{Spectra and data cubes are only available at the CDS via anonymous ftp to cdsarc.u-strasbg.fr (130.79.128.5) or via http://cdsarc.u-strasbg.fr/viz-bin/qcat?J/A+A/vol/page}
	}

	\author{A.~Feldmeier 
			\inst{\ref{inst1}}$^,$\inst{\ref{inst2}}
		\and N.~Neumayer
			\inst{\ref{inst1}}
		\and A.~Seth
			\inst{\ref{inst8}}	
		\and R.~Sch{\"o}del
			\inst{\ref{inst9}}		
		\and N.~L{\"u}tzgendorf
			\inst{\ref{inst12}}
		\and P.~T.~de~Zeeuw
			\inst{\ref{inst1}}$^,$\inst{\ref{inst7}}
		\and M.~Kissler-Patig
			\inst{\ref{inst3}}
		\and S.~Nishiyama
			\inst{\ref{inst10}}
		\and C.~J.~Walcher	
			\inst{\ref{inst11}}	
						}

 \institute{European Southern Observatory (ESO), Karl-Schwarzschild-Stra{\ss}e 2, 85748 Garching, Germany   
 	\label{inst1}
	\and
	\email{afeldmei@eso.org}
		\label{inst2}
          \and
           Department of Physics and Astronomy, University of Utah, Salt Lake City, UT 84112, USA
          	 \label{inst8}
    	 \and
         Instituto de Astrof\'{i}sica de Andaluc\'{i}a (CSIC), Glorieta de la Astronom\'{i}a s/n, 18008 Granada, Spain           	
         \label{inst9}
	          \and 
      ESTEC, Keplerlaan 1, 2201 AZ Noordwijk, The Netherlands 
      \label{inst12}
      	      \and
          Sterrewacht Leiden, Leiden University, Postbus 9513, 2300 RA Leiden, The Netherlands 
          	\label{inst7}
      \and 
          Gemini Observatory, 670 N. A'ohoku Place, Hilo, Hawaii, 96720, USA 
          	\label{inst3}	
	\and
	National Astronomical Observatory of Japan, Mitaka, Tokyo, 181-8588 Japan
		\label{inst10}
	 \and
	 Leibniz-Institut f{\"u}r Astrophysik Potsdam (AIP), An der Sternwarte 16, 14482 Potsdam, Germany
	 	\label{inst11}
          }

\date{Received  March 7, 2014; accepted June 10, 2014}

 \abstract
   {Within the central 10\,pc of our Galaxy lies a dense cluster of stars. This nuclear star cluster forms a distinct component of the Galaxy, and similar nuclear star clusters are found in most nearby spiral and elliptical galaxies.   Studying the structure and kinematics of nuclear star clusters reveals the history of mass accretion and growth of galaxy nuclei and central massive black holes.
}
   {Because the Milky Way nuclear star cluster is at a distance of only 8\,kpc, we can spatially resolve the cluster on sub-parsec scales.  This makes the Milky Way nuclear star cluster a reference object for understanding the formation of all nuclear star clusters. 
   }
   {We have used the near-infrared long-slit spectrograph ISAAC (VLT) in a  drift-scan  to construct  an integral-field  spectroscopic map of the central   $\sim$9.5$\,\!\times\,\!$8 pc  of our Galaxy, and six smaller fields out to 19\,pc along the Galactic plane.
   We use this spectroscopic data set  to extract stellar kinematics both of individual stars and from the unresolved integrated light spectrum.  We present a velocity and dispersion map from the integrated light spectra and model these kinematics using kinemetry and axisymmetric Jeans models.  We also measure radial velocities and CO bandhead strengths of  1,375 spectra from individual stars.
  } 
   {
   We find kinematic complexity in the nuclear star clusters radial velocity map including a misalignment of the kinematic  position angle by 9\degr\,counterclockwise relative to the Galactic plane, and indications for  a rotating substructure perpendicular to the Galactic plane at a radius of 20\arcsec\,or $\sim$0.8pc.
 We determine the mass of the nuclear star cluster within $r$ = 4.2\,pc to (1.4$^{+0.6}_{-0.7}$)\,$\times$~10$^7$~M${_\odot}$. 
  We also show that our kinematic data results in a significant underestimation of the supermassive black hole (SMBH) mass.
  }
{The kinematic substructure and position angle misalignment may hint at distinct accretion events. This indicates that  the Milky Way nuclear star cluster grew at least partly by the mergers of massive star clusters. Compared to other nuclear star clusters,  the Milky Way nuclear star cluster is on the compact side of  the $r_{eff} - M_{NSC}$ relation. The underestimation of the SMBH mass might  be caused by the kinematic misalignment and a stellar population gradient. But it is also possible that there is a  bias    in SMBH mass measurements obtained with integrated light, and this might affect SMBH mass determinations of other galaxies.  
 }
{}
\titlerunning{Large scale kinematics of the Milky Way Nuclear Star Cluster}
\authorrunning{A. Feldmeier et al.}
  \maketitle

\section{Introduction}

The Milky Way nuclear star cluster lies within the central  10\,pc of our Galaxy and is composed of a dense population of stars. The cluster  forms a distinct component, with a half-light radius or effective radius $r_{eff}$  of $\sim$110$-$127\arcsec\,\citep[4.2$-$5\,pc,][]{sb,Fritz14},  and a  mass of (2$-$3)~$\times$~10$^7$~M${_\odot}$ \citep{launhardt02,sb}. However, the formation and growth of the nuclear star cluster and the supermassive black hole (SMBH) in the centre are not  understood.  The stars in the cluster provide a record of the nuclear accretion history and formation processes.

The nuclear star cluster in the Galactic centre is not a unique object, as such clusters  are  common in other galaxies as well. They have been detected in $\sim$\:\!75\% of spiral galaxies \citep{carollo98,boker02}, and spheroidal galaxies \citep{cote06}. These are lower limits, as the presence of bulges and dust lanes can obscure the nuclear cluster and prevent detection  \citep{carollo02,anil06}. 
Galaxies with higher mass than $\sim$\:\!10$^{10}$~M${_\odot}$ usually only host a SMBH, while  nuclear star clusters are preferentially detected in galaxies with lower mass \citep{ferrarese06,wehner06,boeker10,scott13}. But there are also cases where both nuclear star cluster and SMBH coexist in the same galactic nucleus \citep{anil08}. The most convincing case is our own Galaxy.
Precise measurements of stellar orbits around the central radio source Sagittarius A* (Sgr~A*) provide the most direct evidence for the presence of a  SMBH at the centre of our Galaxy. \cite{ghez08} and \cite{gillessen09}  observed a full Keplerian orbit of one of the innermost stars and measured the mass of the black hole as $\sim$\:\!4~$\times$~10$^6$~M${_\odot}$. Also other nuclear clusters show evidence for a central massive black hole from dynamical modelling or the presence of an AGN \citep[e.g.][]{filippenko03,graham09,anil10404,nadine_ncbh12}.

With a distance of only $\sim$\,8\,kpc \citep{ghez08,gillessen09} it is possible to resolve single stars in the Milky Way nuclear star cluster spectroscopically and infer their age. The stellar populations of the central 1\,pc ($\sim$\,26\arcsec) of the  Milky Way  can be studied  only in the near-infrared  (e.g by \citealt{pfuhl11,do13}), as  high extinction makes the Galactic centre impenetrable for observations in the visible bands. The stars are predominantly cool and old  (\textgreater~5\,Gyr old, e.g.  \citealt{blum03,pfuhl11}), but in the central $\sim$\:\!0.5\,pc there exists an additional stellar component in the form of  hot, young stars (4$-$8\,Myr  e.g. \citealt{paumard06,lu09,bartko09}), and stars of intermediate age ($\sim$\,100\,Myr, e.g. \citealt{krabbe_sfh,pfuhl11}).  
 So far, multiple stellar populations have been found in all other nuclear star clusters as well  \citep[e.g.][]{jakob06,rossa06,anil06,siegel07,anil10404,lauerm31_12}, with an underlying old population (\textgreater~1\,Gyr) and a generation of younger stars (\textless~100\,Myr). This suggests a complex star formation history.

There are two prevalent formation scenarios for nuclear star clusters: 1) Stars form in dense clusters elsewhere in the galaxy and then migrate to the galaxy's centre. \cite{tremaine75} suggested the formation of the M31 nucleus by  the infall of globular clusters due to dynamical friction. The clusters would merge to become the observed nuclear star cluster.  
\cite{capuzzo08}, \cite{antonini13},  \cite{gnedin13}, and \cite{antonini14} studied the infall  of  massive clusters in central regions of galaxies  and found that the expected density and velocity-dispersion profiles of the merged  nuclear cluster matches the observations.
2) Nuclear star clusters form in-situ  by gas infall from the disk followed by star formation \citep{milos04,pflamm09}. This theory is supported by the discovery of rotation in the nuclear star clusters of the Milky Way \citep{trippe08,Rainerpm09} and NGC 4244 \citep{anilon4244_08}. A combination of both formation scenarios is also possible \citep{hartmann11,nadine11,turner12,delorenzi13}.

At the Galactic centre the presence of young stars indicates star formation within the past few Myr, as well as the presence of molecular gas within a few pc of Sgr A*.  This demonstrates that the
  necessary material for star formation can be found within the nuclear star cluster. 
The  ring of clumpy gas and dust is called  circumnuclear disk  or circumnuclear ring  and extends to a Galactocentric radius of $\sim$\:\!7\,pc (3\arcmin, e.g. \citealt{yusefh2,christopherhcn,Lee08,baobab12}).  \cite{oka_cnd} suggested that the circumnuclear disk was formed from an infalling disrupted giant molecular cloud, and may eventually  fragment and trigger star formation. \cite{cnd_sf} detected methanol masers, which could hint to an early stage of star formation.

There are several kinematic studies of the stars in the central parsec  of the  Milky Way nuclear star cluster. For example \cite{trippe08} and \cite{Rainerpm09} studied both radial velocities and proper motions but not beyond a distance of 1\,pc from the Galactic centre. They found that the velocity dispersions of the stars are consistent with  an isotropic, rotating cluster. However, the  larger-scale kinematics are not so extensively studied. \cite{Lindqvist921} collected a sample of 134 OH/IR stars at a distance of 5$-$100\,pc ($\sim$\,\!2$-$40\arcmin) from the Galactic centre, but only 15 of their targets are within 8\,pc  from the Galactic centre, and therefore likely cluster members. \cite{mcginn89} obtained integrated light spectra of selected fields with a 20\arcsec\,(0.78\,pc) aperture  at 2.3\,$\mu m$ out to $\sim$\:\!4\,pc (1.7\arcmin) distance along the Galactic plane and $\sim$\:\!1.5\,pc (0.6\arcmin) perpendicular to it. 
They found decreasing velocity dispersion  and increasing velocities with Galactocentric distance, and a flattening of the rotation curve between 2\,pc and 3\,pc ($\sim$\,\!50$-$80\arcsec). But  \cite{mcginn89}   had large scattering in their data.

Hence the Milky Way nuclear star cluster kinematics are known in detail at small scales, but the large scale kinematics beyond 1\,pc remain uncertain. For example, the rotation  is not yet well  determined, although a rotation law can provide insights on the processes that play a role in the formation of the nuclear star cluster.  Calculations of  the Galactic potential beyond 1\,pc   used $\lesssim$\,200 stars to trace the kinematics, but the stars  were spread over a large area of several tens of parsec$^2$. Therefore there is large uncertainty in these measurements. 

In order to overcome this lack of knowledge we have obtained a new spectroscopic data set of the Milky Way nuclear star cluster on a large scale, covering the central $\sim$\:\!9.5\,pc~$\times$~8\,pc  (4\arcmin~$\times$~3.5\arcmin). The purpose of this work is two-fold: (i) to perform the first detailed kinematic analysis of the Milky Way nuclear star cluster on large scales in integrated light; (ii) to construct a dynamical model, in which we derive the cluster mass and constrain the central Galactic potential. 
As by-product we also  extract an H$_2$ gas kinematic map   and almost 1,400 spectra of bright stars. We compute the velocities and CO indices of these stars, from which we can identify young star candidates.

This paper is organised as follows: The observations are summarised in Section~\ref{sec:obs}, and in Section~\ref{sec:datared} we describe the data reduction and analysis. Our results  for the stellar kinematics and stellar populations are presented in Section~\ref{sec:results}. In Section~\ref{sec:sbjeans} we fit  a surface brightness profile and present the results of dynamical Jeans models. We discuss our results in Section~\ref{sec:summary}. Our conclusions are provided in Section~\ref{sec:conclusions}. The H$_2$ gas emission line results are shown in  Appendix \ref{sec:h2}.
Throughout this paper we assume a Galactocentric distance $R_0$ of 8.0\,kpc \citep{distance8kpc}, i.e. 1\arcsec\,corresponds to $\sim$0.039\,pc.

\section{Observations}
\label{sec:obs}
We used the near-infrared long-slit spectrograph ISAAC (Infrared Spectrometer And Array Camera, \citealt{isaac}) in a drift-scan technique to observe the central   $\sim$\:\!9.5\,pc~$\times$~8\,pc  (4\arcmin~$\times$~3.5\arcmin) of our Galaxy. To complement our data set we also observed six fields  out to a distance of $\sim$\:\!19\,pc (8\arcmin) along the Galactic plane, a region where the nuclear stellar disk becomes important. 

\begin{figure}
	\resizebox{\hsize}{!}{\includegraphics{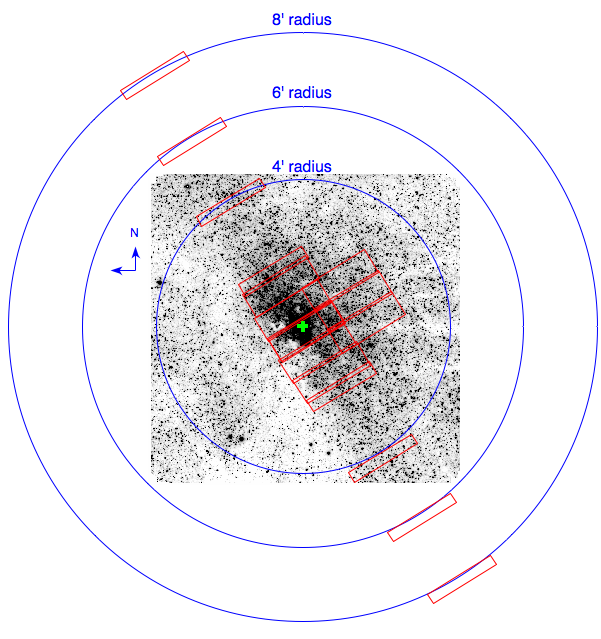} }
	\caption{Positions of the regions scanned with ISAAC (red rectangles). The underlying image is from IRSF/SIRIUS in K$_S$ band \citep{shogo06}. The circles denote a distance of 4\arcmin, 6\arcmin, and 8\arcmin\, from the position of Sgr~A*, respectively,  and the green cross is the position of Sgr~A*. The outer fields are not exactly symmetric due to the drift towards the south-west during the acquisition.}
	\label{fig:region}
	\end{figure}

Observations were performed on VLT-UT3 (Melipal) 
in the  nights of July 3 and 4, 2012. We used the ISAAC short wavelength  medium resolution  spectroscopic mode.  In this mode the spectral coverage is only 0.122$~\mu m$. The central wavelength of our observations is  2.35$~\mu m$, and the resulting spectral range is  $\sim$\:\!2.289\:\!$-$\:\!2.411$\,\mu m$. 
The ISAAC slit has a length of 120\arcsec, and we chose a slit width of 0\farcs6 to obtain a spectral resolution of $R\,=\,4400$, corresponding to $\Delta v\:\!\sim\:\!68\,$km/s. The spatial pixel scale is 0.148\arcsec/pixel, the spectral pixel size is 1.2$~\times~10^{-4}~\mu m$/pixel. The drift scan worked as follows: To cover the Galactic nuclear star cluster to the effective radius, we scanned the slit across the Galactic centre while integrating. During every 120\,s exposure, the slit moved perpendicular to the slit length over 2\arcsec. The scan direction was  along the Galactic plane applying a position angle of 31\fdg40 east of north (J2000.0 coordinates,  \citealt{2004reid}). During the first night we scanned the central $\sim$\:\!120\arcsec\:\!$\times$ 240\arcsec. In the second night we scanned a smaller region north-west of  the centre, which partly overlaps with the observations of the first night, covering $\sim$\:\!120\arcsec\:\!$\times$ 120\arcsec. Additionally we observed six smaller regions with  $\sim$\:\!120\arcsec\:\!$\times$ 16\arcsec\, each along the Galactic plane. They are at distances from the centre of 4\arcmin, 6\arcmin, and 8\arcmin, respectively. The scanned regions are illustrated in Figure~\ref{fig:region}. 
	
For sky observations we made offsets to a dark cloud that is located about 30\arcmin\,east of the Galactic Centre ($\alpha \approx$ 267.00\degr, $\delta \approx$ -28.99\degr,  diameter $\sim$ 11\arcmin, \citealt{dutra_darkcloud_01}).  The sky was also observed in a drift scan with the same scanning velocity.   After a sequence of 8$-$10 object frames we made sky offsets and followed an object - sky - object sequence. We obtained five spectroscopic sky exposures in each sky offset, in order to remove   stars in the sky field and compute a Mastersky frame. Moreover, we observed four B dwarfs as telluric standards in both nights using a standard nodding technique.   For wavelength calibrations we obtained Xenon and Argon arc lamp calibration frames.

\section{Data reduction and Analysis}
\label{sec:datared}
\subsection{Data reduction}
The data reduction of the spectra  includes the following steps: First we use the ISAAC pipeline \emph{isaacp} \citep{isaacdatared} to remove  electrical ghosts from all frames with  the recipe \emph{ghost}, and with the recipe \emph{sp\_ flat} 
we obtain master flats. 
We further  combine the dark files to master darks and subtract them using IRAF (Imaging Reduction and Analysis Facility\footnote{IRAF is distributed by the National Optical Astronomy Observatory, which is operated by the Association of Universities for Research in Astronomy (AURA) under cooperative agreement with the National Science Foundation.}), and remove bad pixels. For cosmic ray removal we use the \emph{Laplacian Cosmic Ray Identification L.A.Cosmic} written by \cite{lacosmic}. Flat fielding as well as distortion correction and wavelength calibrations are performed with  IRAF.
The spectra contain thermal background, which is partly due to the sky and partly to the detector. We make a robust  two-dimensional polynomial fit to each two-dimensional spectral frame along the dispersion axis and subtract the polynomial fit  from the spectra.

Unfortunately, the spectra suffer from persistence. This means that  bright stars, especially those which were saturated in the  images taken before the spectra, are burnt in the detector and remain visible for some time as bright spots in the spectra taken afterwards. Our approach to remove the persistence from the spectra is described in Appendix \ref{sec:pers}. 
  After subtracting the persistence from object and sky spectra we combine the sky frames to master sky frames and  perform sky subtraction with an IDL routine written by  \cite{ohsky}. 

We reduce the telluric spectra using the  double sky subtraction technique, which is the standard for nodding, to ensure optimal sky subtraction.  A telluric correction is performed for every single row of the two-dimensional spectral frames separately with the IRAF task \emph{telluric}. We scale the intensity of the telluric spectra with the difference in  air mass and search for an optimum scale factor, but omit a shift option for the telluric correction. The next step is to shift the reduced spectra  to the local standard of rest using the IRAF recipes \emph{rvcorrect} and \emph{dopcor}.

To determine the astrometry of our spectra, we create images from  spectra taken subsequently before a sky offset.  
We sum the flux of every spectrum along the dispersion axis and reconstruct images, where every exposure extends over 15 columns of the ISAAC pixel scale of 0.148\arcsec/pixel. This means we have a spatial resolution of  2.22\arcsec\,in the drift direction (Galactic east-west) and   a spatial resolution of 0.148\arcsec/pixel along the slit  (Galactic north-south direction).  Every red box in Figure~\ref{fig:region} corresponds to one reconstructed  image. The image is blurred with a Gaussian point spread function (PSF) with the full width half maximum (FWHM) of the seeing during the observations. This  makes the stars less rectangular. The images are  cross correlated with the ISAAC acquisition images that were matched to the 2MASS point source catalogue \citep{2mass} and also blurred. 
While the wavelength range of the acquisition images covers the entire $K$-band, the spectra contain a much narrower wavelength range. Still, we can identify enough stars to perform a meaningful cross correlation. Thus we obtain the astrometry for every spectrum.

We compare the reconstructed image in the new coordinate system with  stars from star catalogues (2MASS point source catalogue by \citealt{2mass}, and the star catalogue constructed by \citealt{shogo06}). Thereby we find a systematic distortion along the slit direction by up to 6 pixels. We measure the positions of bright stars along the slit and compute the offset from the expected position as in  the star catalogue. This deviation is modelled by a second order polynomial, and the data is resampled along the slit direction. After applying the distortion correction to all spectra, the mean offset of the position of bright stars along the slit  to the position of the  star catalogue is reduced from 0.9 to 0.2 pixels, and we rerun the cross correlation. 

We can identify single bright stars in the unbinned data using  the aforementioned  catalogues. 
We select all  stars  brighter than K$_S$ $=$ 11.5$^m$ within our observations and extract their spectra. These  stars are much  brighter than the sun and thus mostly RGB and AGB stars. For comparison, a solar mass star has a K band magnitude of 20$-$21$^m$  at  a distance of 8\,kpc with an extinction of $A_{Ks}$ = 2$-$3$^{m}$.
The position of the star along the slit and its FWHM is fitted with a Gaussian function. We extract the spectrum within   two times the  FWHM\footnote{These spectra are publicly available at the CDS.}.
 The median signal-to-noise ratio (S/N) for these spectra is 25, which is high enough to fit the kinematics.

\begin{figure*}
\centering
\includegraphics[width=18cm]{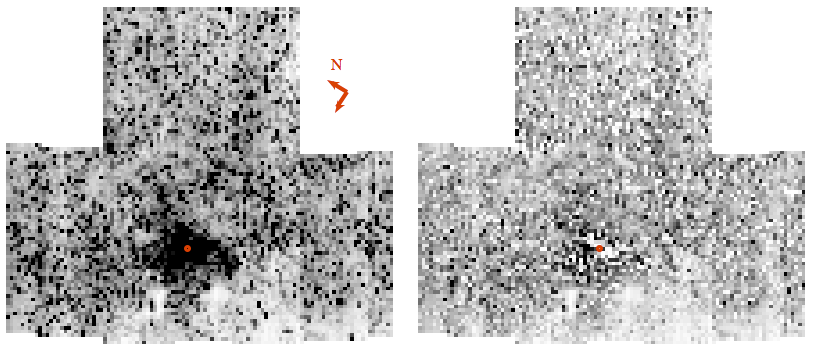}
\caption{Reconstructed data cube from  all central exposures, covering $\sim$\:\!240\arcsec$\times$ 210\arcsec, flux summed over 2.29$-$2.365$~\mu m$. The left panel uses all available data, i.e. the full data cube. The right panel is the data cube after removing foreground and background stars based on their colour, and stars brighter than K=11.5$^m$, i.e. the cleaned data cube. It contains only light from faint member stars. The red point denotes Sgr~A*. The maps are oriented with the Galactic plane  horizontally  and   Galactic North up and Galactic East  left. The arrows indicate the orientation in the equatorial system. Both images are in the same linear flux scaling.}
\label{fig:cube}
\end{figure*}

To examine also the contribution of the fainter stars we construct a single data cube by combining the spectra of the central field  in the form of a ``T'', which is roughly 240\arcsec\,\!$\times$\,210\arcsec (see Figures \ref{fig:region} and \ref{fig:cube}). 
We determine  a relative flux calibration factor for every sequence of subsequently taken spectra  before putting the data set of the central field together. To calibrate the relative flux we use an IRSF/SIRIUS image in K$_S$ band \citep{shogo06}, and the reconstructed images   with the flux  between 2.29 and 2.365$~\mu m$ summed up. 
We estimate the uncertainty for the relative flux calibration to be about 10\%.

To bin the data to one data cube,  we sum up 15 ISAAC pixel rows along the slit, multiply with the  respective flux calibration factor, and obtain a final pixel size of 2\farcs22 $\times$ 2\farcs22. The cube extends over $\sim$\:\!240\arcsec\, along the Galactic plane and $\sim$\:\!210\arcsec\, in perpendicular direction. 
We use our ISAAC data to produce two separate data cubes:
1) A data cube with the full integrated light in each 2\farcs22 $\times$ 2\farcs22 pixel\footnote{This data cube is publicly available at the CDS.}. 
2) A data cube with the brightest stars,  foreground, and background stars removed.
We will refer to these products  as ``full data cube" and ``cleaned data cube".
The full data cube,  integrated in wavelength direction, is shown on the left panel in Figure~\ref{fig:cube}.  For the very centre we have multiple exposures, which are combined by taking the arithmetic mean. We also construct a noise cube, containing the uncertainty of  the persistence removal and sky subtraction.

To produce the cleaned data cube, we cut out stars from the unbinned data before setting up the $\sim$\:\!240\arcsec\,\!$\times$ 210\arcsec\, data cube. 
By cutting out the brightest stars ( K$_S$ $\le$ 11.5$^m$, i.e. K$_{cut}$=11.5$^m$), we  obtain light just from the underlying fainter population, reducing the effects of  shot noise caused by single stars in our data cube. In addition to removing the brightest stars, we  also use colour information to remove the  fainter foreground and background stars. 
For this purpose we apply colour cuts and assume that all stars with H$-$K$_S$ \textless~1.5$^m$ or H$-$K$_S$ \textgreater~3.5$^m$ are foreground or background stars, respectively.  Our exclusion criterion is more inclusive than the one applied by \cite{distance}, who excluded all stars with H$-$K$_S$ \textless~1.8$^m$ or H$-$K$_S$ \textgreater~2.8$^m$. We remove foreground and background stars brighter than magnitude K$_S$ = 14$^m$. For stars fainter than K$_S$ = 14$^m$ we cannot reliably determine their position along the slit and might misidentify a cluster member star. 
After fitting the position and FWHM of a star which we want to cut out,  we set the pixel counts within one FWHM on either side of the position of the star to zero. 
The integrated light of the cleaned data cube is shown in Figure~\ref{fig:cube} (right panel). Some of the pixels are empty and hence contain no information at all.

\subsection{Deriving stellar kinematics}
\label{sec:co}
 To derive stellar kinematics we fit the line-of-sight-velocity distribution (LOSVD) of the CO absorption lines (2.2902$-$2.365$~\mu m$).
These lines are most prominent in cool late-type stars with  several Gyr age. 
Moreover,   the CO lines are an excellent tracer of the stellar kinematics in the highly extinction affected Galactic centre region because the CO lines lie within the near-infrared $K$-band, where extinction is low enough to allow for sensitive observations. To obtain the LOSVD, we use the IDL routine \emph{ppxf} \citep{ppxf}. 
It recovers the Gauss-Hermite parameters ($V, \sigma, h_3, h_4, ...$) of the LOSVD by convolving template spectra with the parameterised LOSVD and finding the best fit to the observed spectrum in pixel space. We fit only the first two moments of the LOSVD, the first moment  corresponds to the velocity, the second moment to the velocity dispersion.  As template spectra we use the high-resolution (R~$\ge$~45,000) spectra of \cite{wallace}, which contain supergiant, giant,  and dwarf star spectra in the  spectral types from G to M. 
Those spectra are convolved with a Gaussian to obtain the same spectral resolution as our data set. The program \emph{ppxf} finds an optimal template, which is a linear combination of the template spectra.  The optimal template convolved with the LOSVD recovers the shape of the integrated light spectrum.
 Errors are calculated from  Monte Carlo simulations.  We add random noise to the spectra in each pixel and calculate the LOSVD  in 100 runs. The fits are performed in the wavelength range of 2.2902$-$2.365\,$\mu m$\,, which contains four CO absorption lines, $^{12}$CO~(2-0), $^{12}$CO~(3-1),  $^{13}$CO~(2-0), and $^{12}$CO~(4-2).  
 
\begin{figure}
\resizebox{\hsize}{!}{\includegraphics{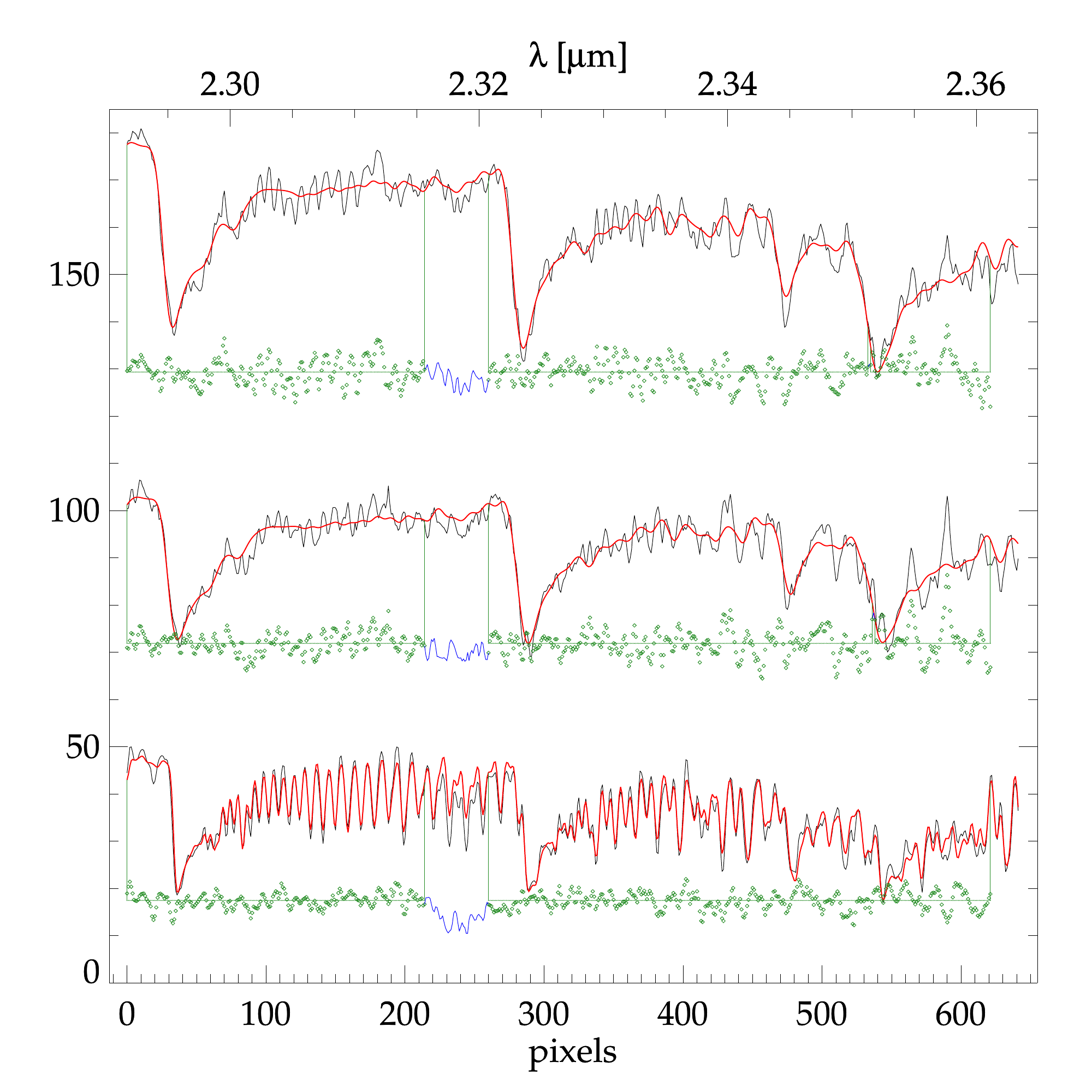} }
\caption{Example of CO absorption  line spectra (black) of a Voronoi bin of the full data cube (upper spectrum), a Voronoi bin of the cleaned data cube (middle spectrum) and of a single star (lower spectrum). The red lines are the \emph{ppxf}-fits to the data. The blue regions are not fitted, as there is  a strong telluric absorption line in this wavelength region.  Green dots are the residuals of the fits. The fluxes of the spectra are scaled for this plot.}
\label{fig:cofit}
\end{figure}

\begin{figure*}
\centering
\includegraphics[width=18cm]{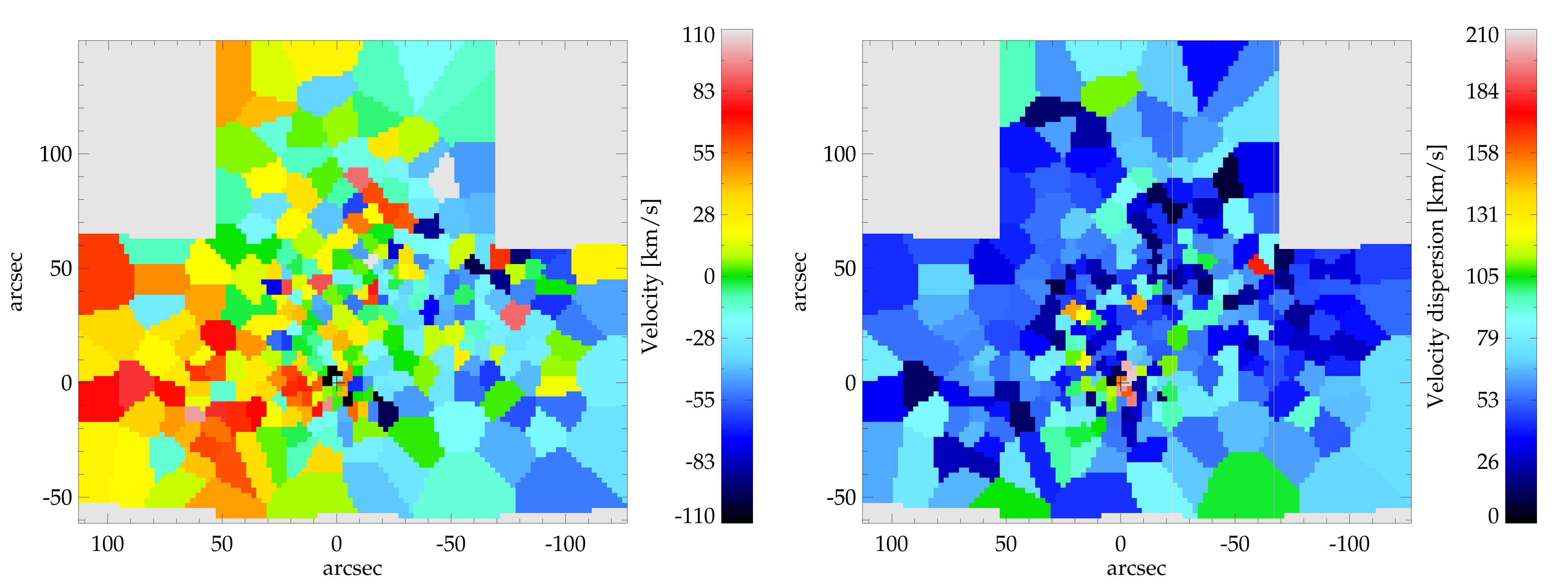} 
\caption{CO absorption  line map of  velocity, and velocity dispersion (left, and right, respectively) obtained by using the full data cube.  Both, velocity and velocity dispersion are in units of km/s. The coordinates are centred on Sgr~A* and along the Galactic plane with a position angle  of 31\fdg40. The  plus sign marks the position of Sgr  A*.  
The velocity is in the local standard of rest, Galactic North is up, Galactic East is left.}
\label{fig:map60}
\end{figure*}

\begin{figure*}
\centering
\includegraphics[width=18cm]{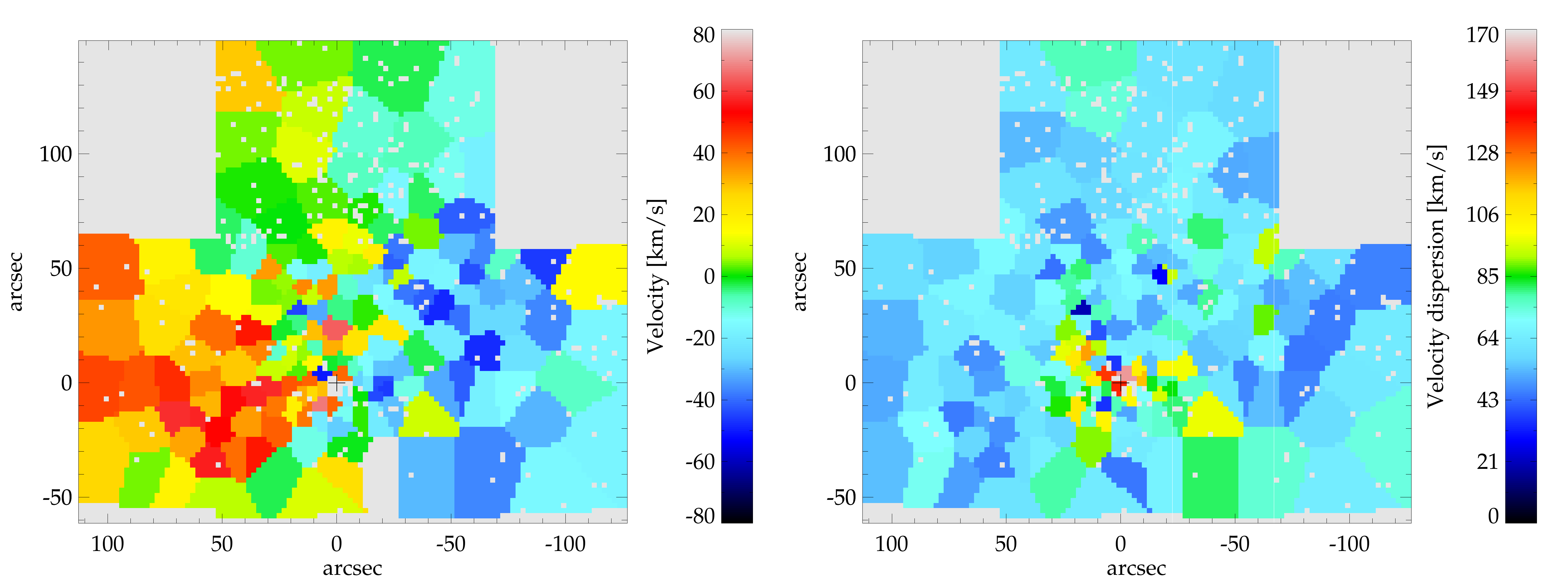}
\caption{Same as Figure~\ref{fig:map60}, but  for the  cleaned data cube. White pixels mark regions where there was no signal left after cutting out stars. }
\label{fig:map_mag60}
\end{figure*}

We use the two data cubes to  obtain a velocity and velocity dispersion map of the integrated light. 
By applying adaptive spatial binning  \citep{voronoi} to our data we  make sure that the integrated spectrum of every bin has approximately the same  signal-to-noise ratio (S/N). Therefore we calculate the S/N for every single pixel of the data cubes with \emph{ppxf} before the binning. We then sum the spectra of all pixels in a Voronoi bin to one integrated light spectrum per bin and fit the LOSVD. Figure~\ref{fig:cofit} shows three example spectra, and the respective fits. The upper spectrum is an integrated light spectrum from the full data cube, the middle spectrum is from the cleaned data cube, and the lower spectrum is  from a single bright star.  Figure~\ref{fig:map60} shows the velocity and velocity dispersion maps for a binning of at least S/N\,\!=\,\!60 for the full data cube.  One can  see the rotation of the  nuclear star cluster around Sgr~A*, and an increase of the velocity dispersion up to $\sim$\,\!200\,km/s. The bright supergiant star IRS~7  with  apparent magnitude K\,\!$\sim$\,\!7.0$^m$  \citep{holographie}  dominates its bin, and it has therefore only low velocity dispersion. 
Figure~\ref{fig:map_mag60} shows the kinematic maps for the data cube where single stars are cut out. It appears smoother and less affected by shot noise. Some pixels contain no information at all after the bright stars and foreground stars are cut out, and those pixels are displayed in white.

To quantify the amount of shot noise in the cleaned data cube we produce  two additional cleaned data cubes for which we vary the magnitude of the stars we cut out from K$_{cut}$~=~11.5$^m$ to  K$_{cut}$~=~11$^m$ and K$_{cut}$~=~12$^m$. 
On these data cubes we apply the same Voronoi binning as on the cleaned data cube. The number of stars per bin in the data cube with K$_{cut}$ = 11$^m$ increases by 1.85 on average, and decreases in the data cube with K$_{cut}$ = 12$^m$ by 2.3 on average.
We find that a variation of the magnitude cut by 0.5$^m$ has no strong influence on our results. In 75\% of all Voronoi bins the difference of the velocity measurements by varying K$_{cut}$ by 0.5$^m$ is less than the velocity uncertainties. For only  5\% of all Voronoi bins the difference is more than two times the velocity uncertainties. The difference of the velocity measurements has a standard deviation of 8.2\,km/s.  There  is also no systematic trend in the velocity dispersions obtained from the different cleaned data cubes.  The difference of the velocity dispersions in the same bin from the different cleaned data cubes has a mean value of  0.7\,km/s. We conclude that  on average our  measurements are robust and not severely influenced by shot noise.

We further estimate the shot noise in the velocity map obtained from the full data cube by a comparison of the total flux in each Voronoi bin with the flux given  by a $K$-band luminosity function (KLF). Therefore we use the surface brightness profile from Section~\ref{sec:sb} scaled to the $K$-band. We calculate the total flux for each Voronoi bin by multiplying the surface brightness at a given bin with the respective area covered by the bin. The KLF tells us how many stars of a given magnitude are present in the nuclear star cluster. We use the KLF of \citet[their Table~3]{zoccali03}, which was measured in the Galactic bulge. However, \cite{genzel03} showed that the KLF of the central 9\arcsec\,of the Milky Way nuclear star cluster is very similar to the KLF measured by \cite{zoccali03}. For a rough estimation of the number of stars we can therefore assume that a normalised KLF does not vary strongly over the entire field.  We compute the integrated flux of the normalised KLF. Therefore we  consider the KLF only for stars brighter than K=17$^m$, since this is  the approximate magnitude limit for red clump stars, and fainter stars have rather weak CO absorption features.  A comparison with the total flux of the Voronoi bins tells us that in each bin we expect at the least 740 stars and 5,500 stars on average. When we assume that the brightest star in a Voronoi bin has a magnitude of K=7.27$^m$, the fraction of light it would contribute to a Voronoi bin is 12\% on average, and 51\% in the worst case. The high light fraction of single stars indicates that shot noise can have a severe effect on results obtained from  the full data cube. 

 According to the KLF, the stars fainter than K=11.5$^m$   contribute 36\% of the total flux of the nuclear star cluster. So,  in the Voronoi bins of the cleaned data cube the total flux is decreased to 36\%.  We compare the remaining flux of a Voronoi bin of the cleaned data cube with the flux of  a star with K=11.52$^m$. We found that a single star of K=11.52$^m$  contributes at maximum 1.9\% of the remaining light of the Voronoi bin. This is significantly decreased compared to the maximum light fraction of 51\% of the full data cube. Even if all the light of the cleaned data cube would come from the brightest stars with K=11.52$^m$, there would be on average 400 of these stars in each bin, and at the least 53 stars. This shows that the influence of a single star is significantly decreased in the cleaned data cube, and shot noise is no longer an issue.

\begin{figure*}
\centering
\includegraphics[width=18cm]{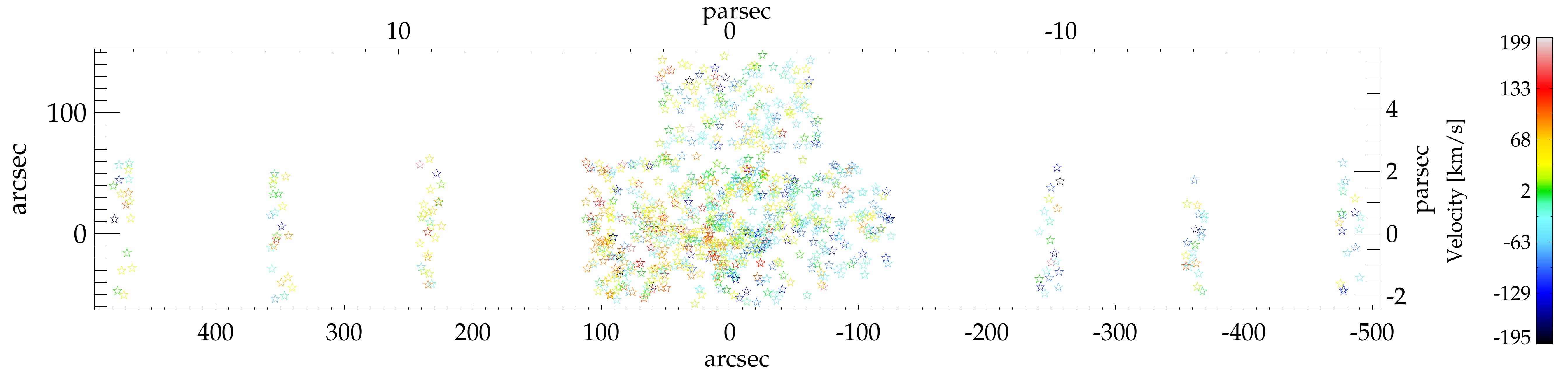}
\caption{Velocity map of 1,094 bright stars (K $\le$ 11.5$^m$) with CO absorption lines in the colour interval 1.5$^m$ $\le$ H$-$K $\le$ 3.5$^m$. We plot only stars with S/N\,\textgreater 12 and $|v|$ \textless\,200\,km/s.}  
\label{fig:map_velbright}
\end{figure*}  

We also fit the LOSVD for each spectrum of the single bright stars with K$_S \le$ 11.5$^m$. From these 1,375 spectra we find that $\sim$1,200 are from stars in the colour range 1.5$^m$ $\le$ H$-$K $\le$ 3.5$^m$, suggesting their membership to the nuclear star cluster. Some of the spectra are from the same star, since we had overlapping exposures. The result for ten of these stars is shown in Table~\ref{tab:sphe}, the full Table is publicly available together with the spectra at the CDS.  Figure~\ref{fig:map_velbright} displays the  velocity map of stars in the colour interval 1.5$^m$ $\le$ H$-$K $\le$ 3.5$^m$. The mean  velocity of our data set is consistent with zero, and the standard deviation is 74\,km/s. The most extreme radial velocities we obtain are -340.3\:\!$\pm$\:\!59.5\,km/s on the blue side, and 291.9\:\!$\pm$\:\!1.5\,km/s on the red side. We also  see rotation in the same sense as the Galactic rotation.

\begin{table*}
\caption{Table of stars with  coordinates RA and Dec, K and H band magnitude according to the source catalogue, measured line-of-sight velocity $V$,  CO index $CO_{mag}$ and signal-to-noise ratio S/N. We only reproduce 13 lines of this Table here. A full Table with the spectra is available in electronic form at the CDS.}
 \label{tab:sphe}
 \centering
\begin{tabular}{lccrrcrcc}
\noalign{\smallskip}
\hline\hline
\noalign{\smallskip}
ID &RA&	Dec&	K\,\,\,\,&	H \,\,\,\,&	source &V\,\,\,\,\,\,\,\,\,\,\, & $CO_{mag}$& S/N\\
&$\left[  \degr \right]$& $\left[  \degr \right]$	 &$\left[  ^{m} \right]$\,\,&$\left[   ^{m} \right]$\,\,\,	&  catalogue &$\left[  km/s \right]  \,\,\,\,$&&\\

\noalign{\smallskip}
\hline
\noalign{\smallskip}
1&$       266.44606$ & $       -28.99290 $&$  11.24$ & $  13.50 $& SIRIUS& $    86.7\,\pm\,     1.6$ & $ 0.31$ & $ 21.1$ \\
2&$       266.44566$ & $       -28.99287 $&$   9.73$ & $  11.28 $& 2MASS & $    80.9\,\pm\,     1.5$ & $ 0.45$ & $ 18.9$ \\
3&$       266.44671$ & $       -28.99027 $&$  11.02$ & $  14.12 $& SIRIUS& $    26.3\,\pm\,     3.6$ & $ 0.40$ & $ 18.7$ \\
4&$       266.44510$ & $       -28.99095 $&$  10.54$ & $  12.13 $& 2MASS & $   -55.5\,\pm\,     0.2$ & $ 0.48$ & $ 22.4$ \\
5&$       266.44680$ & $       -28.98703 $&$  10.82$ & $  12.28 $& 2MASS & $    34.9\,\pm\,     2.5$ & $ 0.39$ & $ 22.9$ \\
6&$       266.44453$ & $       -28.98938 $&$  11.48$ & $  14.13 $& SIRIUS& $    -4.9\,\pm\,     1.2$ & $ 0.26$ & $ 22.9$ \\
7&$       266.44371$ & $       -28.99045 $&$  10.32$ & $  11.91 $& 2MASS & $    77.4\,\pm\,     1.8$ & $ 0.40$ & $ 17.9$ \\
8&$       266.44243$ & $       -28.99068 $&$  11.33$ & $  13.71 $& SIRIUS& $    63.3\,\pm\,     0.3$ & $ 0.39$ & $ 21.9$ \\
9&$       266.44301$ & $       -28.98553 $&$  10.44$ & $  13.40 $& SIRIUS& $    93.0\,\pm\,     1.0$ & $ 0.30$ & $ 16.2$ \\
10&$       266.44122$ & $       -28.98636 $&$   9.34$ & $  11.21 $& 2MASS & $   -38.9\,\pm\,     0.7$ & $ 0.51$ & $ 20.2$ \\
\noalign{\smallskip}
high velocity stars \\
521&$       266.41859$ & $       -29.00958 $&$   8.66$ & $  11.06 $& SIRIUS& $  -340.3\,\pm\,    59.5$  \tablefootmark{a} & $ 0.07$ & $ 16.7$ \\
1042&$       266.37746$ & $       -28.99331 $&$  11.41$ & $  13.35 $& SIRIUS& $   291.9\,\pm\,     1.5$ & $ 0.20$ & $ 22.2$ \\
1056&$       266.39317$ & $       -29.01596 $&$  10.14$ & $  11.87 $& 2MASS & $  -265.6\,\pm\,     0.5$ & $ 0.11$ & $ 45.2$ \\

\noalign{\smallskip}
  \hline
\noalign{\smallskip}
\end{tabular}
\tablefoot{
\tablefoottext{a}{The high uncertainty is caused by the limited wavelength range. In this spectrum there is no continuum on the blue side of the CO absorption line. However, this star is  known as IRS~9 and its   velocity  was  measured by \cite{rv_fast_08} to $-347.8$\,km/s.}
}
\end{table*}

To ensure the accuracy of our wavelength calibration and velocity determination, we compare our derived velocities  to previous stellar velocity measurements.  The cleanest comparison is for the bright supergiant IRS~7, where we find a velocity of $-$116$\pm$1\,km/s, in excellent agreement with the value from \citet[$-$114$\pm$3\,km/s]{maser}.  For other stars our low spatial resolution along the galactic plane creates possible issues with contamination in our data.  Comparing stellar velocities derived from SiO masers by \cite{deguchi}, we find that four stars have $\Delta v \leq 6$\,km/s (stars with names 3-6, 3-16, 3-5.2 and 16-49 in \citealt{deguchi}). Two other stars have a 30\,km/s  and 60\,km/s offset in velocity (stars with names 3-885 and 3-57 in \citealt{deguchi}).  These larger offsets could result from intrinsic changes in the velocities in pulsing AGB stellar atmospheres \citep{agb}, but for star 3-885 a misidentification is also possible.  Overall, it appears that our absolute velocities are accurate to within $\sim$5\,km/s.

The large field of view results in the radial direction   not being constant  over the entire field and the proper motion of the sun with respect to the Galactic centre has to be taken into account. 
We test the amplitude of this so-called perspective rotation using Equation 6 of \cite{ven}. Assuming that the sun moves  with $v_x = $ 220\,km/s in the Galactic plane \citep{speedofsun} at a distance of 8\,kpc from the centre, and  with $v_y = $ 7\,km/s towards the Galactic North, we find that this effect is negligible perpendicular to the Galactic plane. Along the Galactic plane the perspective rotation is $\lesssim$\,0.13\,km/s for the central field. For the outermost fields at 8\arcmin\,distance from the centre, it is 0.5\,km/s. This is less than the usual velocity uncertainty we obtain. Nevertheless, we corrected the velocities of the six outer fields shown in Figure~\ref{fig:map_velbright} for perspective rotation.

\subsection{Line strength measurements}
\label{sec:coindex}
\begin{figure}
\resizebox{\hsize}{!}{\includegraphics{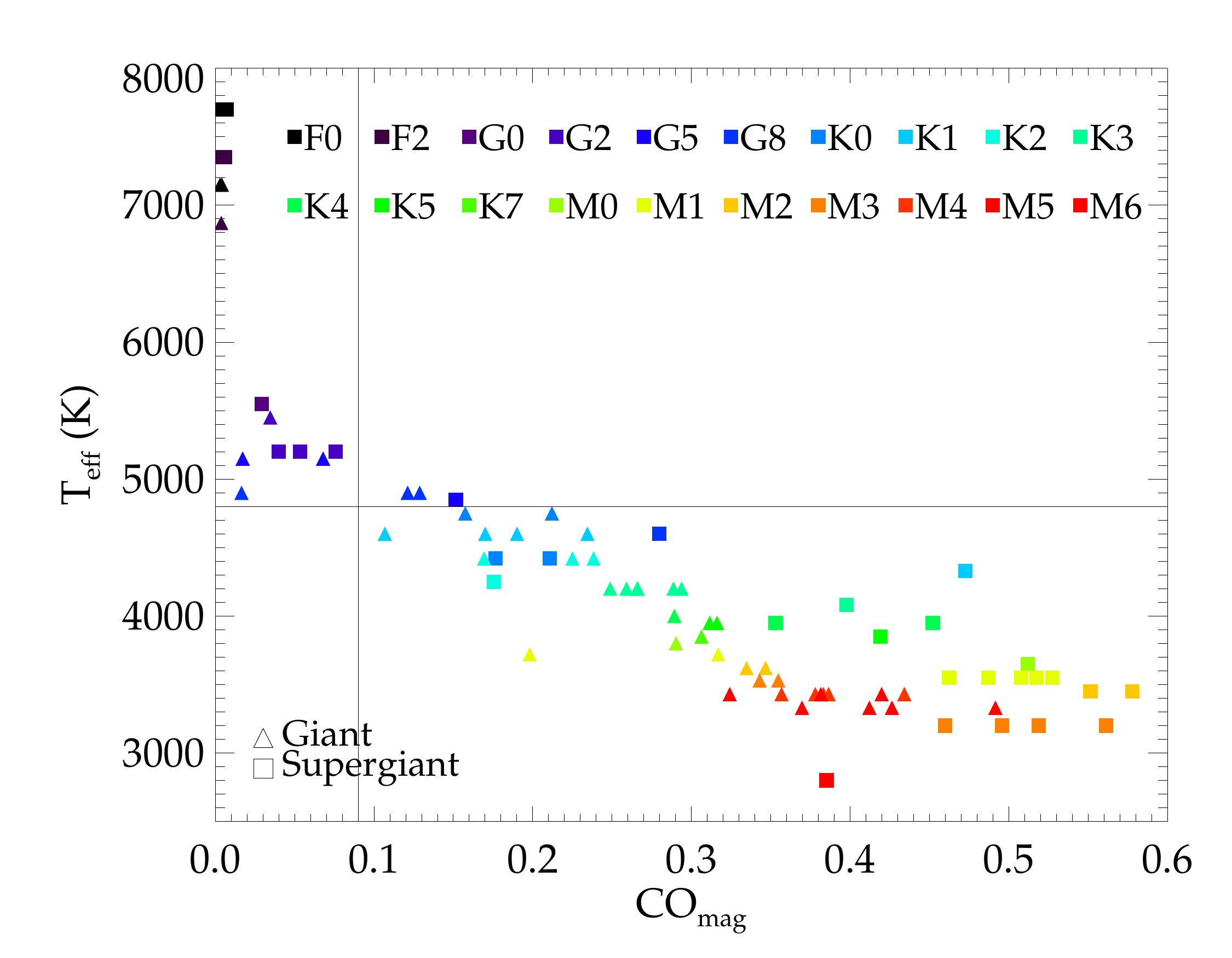}}
\caption{Relationship between effective temperature $T_{eff}$ in $K$ and the CO index $CO_{mag}$ for giant (triangle symbol) and supergiant (square symbol) stars of the IRTF Spectral Library \citep{irtf}. Different colours denote a different spectral type. The black horizontal line marks 4,800\,K, the black vertical line marks $CO_{mag}$=0.09  }
\label{fig:cotemp}
\end{figure}

 With our data set we can give a rough age estimation of bright stars. We use the fact that  CO absorption lines are prominent in old stars, but not in young, hot stars. To quantify the depth of the CO line, we use  the CO index $CO_{mag}$ in a colour-like  way  as defined by \cite{kleinmannhall}
\begin{equation}
CO_{mag}  =  -2.5 \log \frac{F_a}{F_c},
\end{equation}
where 
$F_a$ is the mean flux at the first CO absorption line (2.2931$-$2.2983\,$\mu m$), and $F_c$ is the mean flux at the continuum (2.2885$-$2.2925\,$\mu m$\,), corrected for  radial velocity shifts. This is not the same continuum wavelength region as defined in \cite{kleinmannhall}, which is at  shorter wavelengths. As we are very limited in our spectral range, we have to modify the CO index definition.  To calibrate our $CO_{mag}$ with a temperature, we use the IRTF Spectral Library \citep{irtf}. The metallicities of both the stars in the library and in the Milky Way nuclear star cluster are near-solar \citep{cunha07,pfuhl11}. We compute $CO_{mag}$ for giants and supergiants, with the spectral class ranging from F0 to M6. \cite{spectype} lists the temperature of stars depending on the spectral class and type. The relation between effective temperature $T_{eff}$ and $CO_{mag}$ is shown in Figure~\ref{fig:cotemp}. We found that for stars with a temperature $T$~\textgreater~4,800\,K, the $CO_{mag}$ is less than 0.09.  Using Padova isochrones for solar metallicity stars with $M_K \le -3^m$ (which corresponds to K $\le$ 11.5$^m$ in the Galactic centre), we find that stars hotter than 4,800\;K are younger than $\sim$300\,Myr. 
Therefore we conclude that stars with $CO_{mag}$\,\textless\, 0.09 are younger than 300\,Myr. 
 Also for higher values of $CO_{mag}$, the CO index correlates roughly with  the age of the stars. 

\section{Stellar Kinematics and Population results}
\label{sec:results}

\subsection{Kinematic structure and substructure of the nuclear star cluster}
\label{sec:intlight}

 In this section we show that the overall rotation of the Milky Way nuclear star cluster is misaligned by $\sim$9\degr$\pm$3\degr\,counterclockwise from the major axis of the galaxy at radii from 1 to 4 pc.  Furthermore, we find evidence for  one cold  rotating substructure within the central parsec.

 \begin{figure}
\resizebox{\hsize}{!}{\includegraphics{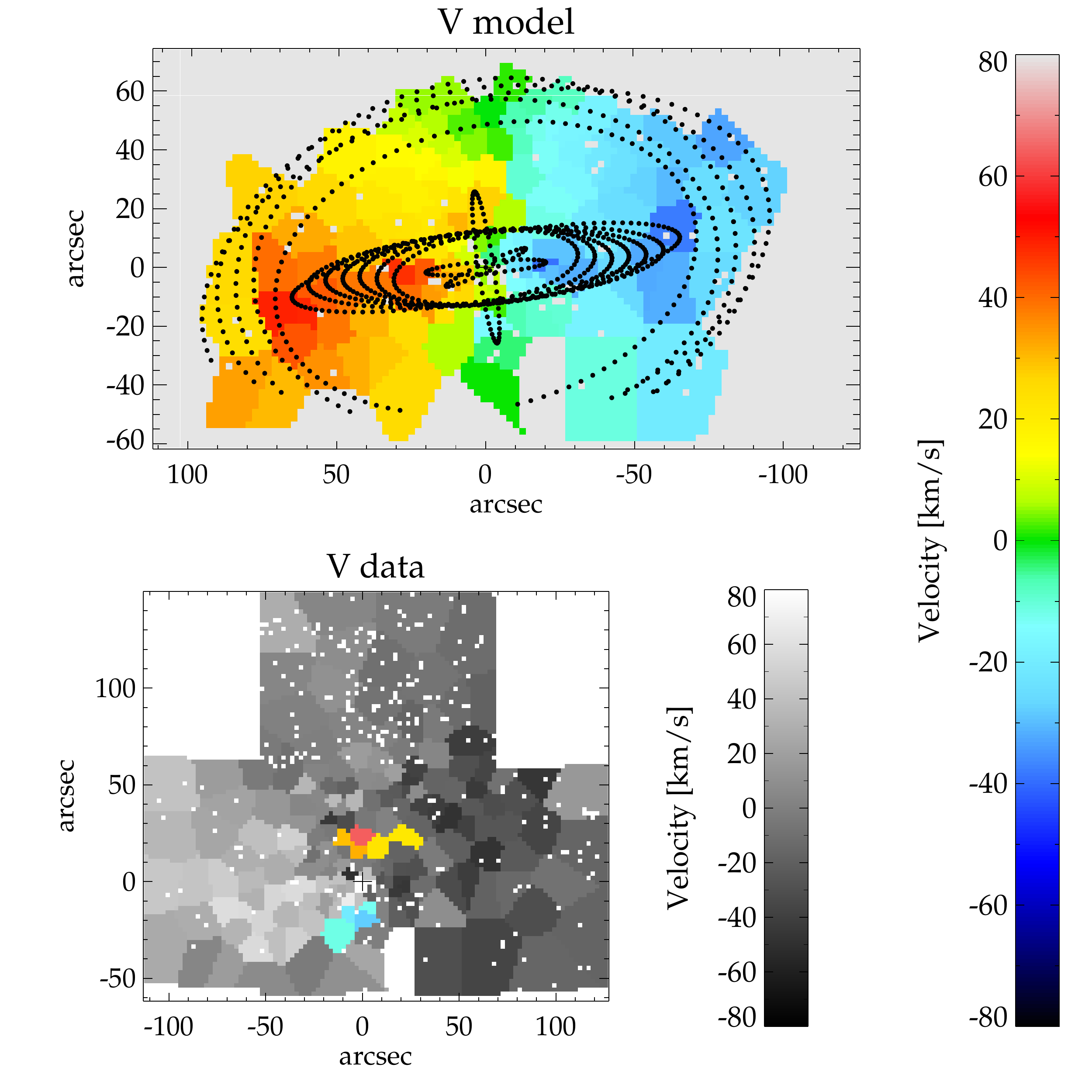}}
\caption{Upper panel: Kinemetric model velocity  map of the cleaned data cube. Black dots denote the best fitting ellipses. The model goes only to $r$ $\sim$ 100\arcsec\, along the Galactic plane and to $\sim$ 60\arcsec\, perpendicular to it. The Voronoi bin with the highest uncertainty was excluded from the model. Lower panel: The velocity map as in Figure~\ref{fig:map_mag60} shown in grayscale, the bins that show rotation perpendicular to the Galactic plane are overplotted in colour scale.  }
\label{fig:kinemetry}
\end{figure}

We  model the stellar kinematics of the velocity maps  using Kinemetry, which was developed by \cite{kinemetry}. Kinemetry assumes that the velocity profile along an ellipse around the centre  can be expressed by a simple cosine law. The IDL routine written by \cite{kinemetry} divides the velocity map into individual elliptical rings, which are described by harmonic terms. The best fitting  ellipse depends on two parameters, the   kinematic position angle PA$_{kin}$, and  kinematic axial ratio $q_{kin}$ ($=1-\epsilon_{kin}$).  By default $q_{kin}$ is constrained to the interval [0.1, 1], while we let PA$_{kin}$ unconstrained.

The result of the kinemetric analysis of the velocity map from the cleaned data cube (Figure~\ref{fig:map_mag60}) is listed in Table~\ref{tab:kinemetry}, and  the upper panel of Figure~\ref{fig:kinemetry} shows the kinemetry model velocity map.  From the axial ratio one can distinguish three families of ellipses. The three innermost ellipses form the first family, the next seven ellipses build the second family, and the outermost five ellipses form the third family.

For the two outer families the kinematic position angle PA$_{kin}$ is 4$-$15\degr\,Galactic east of north, with a median value of 9\degr.  However, the photometric position angle PA$_{phot}$ was measured by \cite{sb} using \emph{Spitzer} data to $\sim$0\degr.  This means that there is an offset between PA$_{phot}$ and PA$_{kin}$.  To test whether this offset  could be caused by extinction from  the 20\,km/s cloud \citep[M-0.13-0.08, e.g.][]{20kmsref} in the Galactic southwest, we flag all bins in the lower right corner as bad pixels and repeat the analysis, but the position angle offset remains. We  test the effect of Voronoi binning by running kinemetry on a velocity map with S/N=80. While the values of $q_{kin}$ for the second family are by up to a a factor two higher with this binning, the PA$_{kin}$ fit is rather robust. We obtain a median value for PA$_{kin}$ of  12.6\degr\,beyond a semi-major axis distance of r~$\sim$~40\arcsec. We conclude that the effect of the binning can vary the value of the PA$_{kin}$, but the PA offset from the Galactic plane is robust to possible dust extinction and binning effects. 
Also our cleaning of bright stars and foreground stars may cause  a bias in the PA$_{kin}$ measurements. For comparison we run the kinemetry  on the velocity map of the full data cube (Figure~\ref{fig:map60}). In this case there is higher scattering in the kinemetric parameters, caused by shot noise. 
However,  beyond 35\arcsec\,semi-major axis distance the median PA$_{kin}$ is at 6.1\degr, i.e. the PA offset is retained. This smaller value could come from the contribution of foreground stars, which are aligned  along the Galactic plane.  It could also mean that  bright stars are not as misaligned to the photometric major axis as fainter stars are. Young stars tend to be brighter, thus the integrated light likely samples an older population than the individual stars. 
Analysis of the resolved stars in the colour interval 1.5$^m$ $\le$  H$-$K $\le$ 3.5$^m$ and in the radial range of 50\arcsec\,to 100\arcsec\,shows an offset in the rotation from the Galactic plane by (2.7$\pm$3.8)\degr.  
As previous studies focused on the brightest stars of the cluster, the PA offset of the old, faint population remained undetected.

In the innermost family there is one ellipse with a position angle of $-$81.5\degr, i.e. PA$_{kin}$ is almost perpendicular to the photometric 
position angle PA$_{phot}\approx 0\degr$. This is caused by a  substructure at $\sim$20\arcsec\,north and south of Sgr~A*, that seems to rotate on an axis perpendicular to the Galactic major axis.  
This feature is highlighted in the lower panel of Figure~\ref{fig:kinemetry}. North of Sgr~A* we find  bins with velocities of 20 to 60\,km/s, while in the Galactic south  bins with  negative velocities around $-10$ to $-30$\,km/s are present.  The feature expands over several Voronoi bins north and south of Sgr~A*. It extends over $\sim$35\arcsec\,(1.4\,pc) along the Galactic plane, and $\sim$30\arcsec\,(1.2\,pc) perpendicular to it.

This substructure also causes the small axial ratio values of the second family of ellipses between 30\arcsec\,and 70\arcsec\,in our kinemetry model. All semi-minor axis distances from Sgr~A* are below 20\arcsec, i.e. at smaller distances to Sgr~A* than the perpendicular substructure. Only  the third family of ellipses, which has semi-major axis values above 70\arcsec, skips over this substructure and reaches   higher values of $q_{kin}$.

 To check if the perpendicular rotating   substructure is real, we apply Voronoi binning with a higher S/N of 80 instead of 60, and obtain again this almost symmetric north-south structure. Also with a lower S/N of 50, the substructure appears in both data cubes.  We also check the influence of the cleaning from bright stars on this feature using the cleaned maps with K$_{cut}$ =11$^m$ and K$_{cut}$ = 12$^m$. The substructure remains also in these data cubes, independent of the applied binning. 
The  fact that this feature persists independent on the applied magnitude cut, or binning, and that it extends over several bins,  indicates that the observed kinematic structure is not caused by shot noise of individual stars.  

We run a Monte Carlo simulation to test if shot noise can mimic the observed substructure. We do this in 1,000 runs by adding random velocities to a smooth velocity map and running Kinemetry on the simulated velocity maps. Then we test if there are substructures in the simulated velocity maps. 
First, we construct a smooth velocity map using Kinemetry on the velocity map of the cleaned data cube. By constraining the PA$_{kin}$ to the interval $\left [-25\degr, 25\degr \right ]$ we obtain a velocity map without any substructures. To this smooth velocity map we add random velocities drawn from a normal distribution with a mean velocity of 0\,km/s, and a standard deviation of $\sigma$ = 8.2\,km/s. The value of $\sigma$\,=\,8.2\,km/s is a measure for the shot noise (see Section~\ref{sec:co}). On  1,000 simulated velocity maps  we run Kinemetry models. Then we analyse if there are  substructures in the simulated velocity maps. A substructure has to fulfil the following criteria: 1.) In  the Kinemety model there is an ellipse with a  PA$_{kin}$ that deviates by more than 45\degr\,from the median of the other PA$_{kin}$  in this run; 2.) The velocity of this ellipse $V_{rot}$ is greater than 8.2\,km/s.  Our Monte Carlo simulation reveals that in  1,000  simulated velocity maps  only 9.1\% have such a substructure. Only 0.2\% have a substructure with $V_{rot}$ \textgreater 20\,km/s, which is comparable to the value of $V_{rot}$  = 24.1\,km/s we find in the ISAAC data (see Table~\ref{tab:kinemetry}). We conclude that the feature is significant at 99.8\% and it is  unlikely that a statistical fluctuation can mimic a kinematic substructure in our data.

We do not find a  velocity substructure 
for the single bright stars  as seen in the integrated light. In both the northern and the southern region, the median velocities are $\sim$3\,km/s, i.e. consistent with zero. 
We note that the bright stars likely represent different stellar populations than the unresolved light. The bright stars are  dominated by younger populations with supergiants and AGB stars, while the unresolved light is dominated by older populations, where supergiants and AGB stars are rare.  Thus this substructure, if real, would likely have an older stellar population.
We discuss this discovery in more detail in Section~\ref{sec:disussionkin}.

The rotation velocity obtained by the kinemetry model is at median values of $\sim$\,\!33\,km/s  for all different tested data cubes and binnings. The exact shape of the curve however depends on the fitted ellipse axial ratios, which varies with the binning of the velocity map.
\begin{table}
\caption{Result of the kinemetric analysis of the cleaned data cube velocity map. The  columns denote the semi-major axis distance $r$ in arcsec of the best fitting ellipses to Sgr~A*,  the kinematic position angle PA$_{kin}$ in degrees, the kinematic axial ratio $q_{kin}$, and the rotation velocity $V_{rot}$ in km/s. The position angle is defined such that 0\degr\,corresponds to rotation around the Galactic North-South direction.}
\label{tab:kinemetry}
\centering
\begin{tabular}{c r r r}
\noalign{\smallskip}
\hline\hline
\noalign{\smallskip}
$r$&$PA_{kin}$\hspace*{2 mm}&$q_{kin}$\hspace*{5 mm}&$V_{rot}$\hspace*{4 mm}\\
$\left[ arcsec \right] $& $\left[ \degr \right]$\hspace*{4 mm}& &$\left[ km/s\right]$ \hspace*{2 mm} \\
\noalign{\smallskip}
\hline
\noalign{\smallskip}
$15.0$ & $  25.2\,\pm\, 1.4 $&$  0.10\,\pm\,  0.00 $&$  30.5\,\pm\,  1.4$ \\
$20.5$ & $   4.9\,\pm\, 2.0 $&$  0.10\,\pm\,  0.00 $&$  43.9\,\pm\,  1.1$ \\
$26.1$ & $ -81.5\,\pm\, 1.3 $&$  0.10\,\pm\,  0.00 $&$  24.1\,\pm\,  1.4$ \\
\noalign{\smallskip}
\hline
\noalign{\smallskip}
$31.7$ & $   9.1\,\pm\, 1.3 $&$  0.37\,\pm\,  0.01 $&$  38.5\,\pm\,  1.1$ \\
$37.3$ & $   6.7\,\pm\, 1.2 $&$  0.34\,\pm\,  0.01 $&$  31.5\,\pm\,  0.9$ \\
$43.1$ & $   4.4\,\pm\, 1.0 $&$  0.29\,\pm\,  0.01 $&$  32.2\,\pm\,  0.7$ \\
$48.9$ & $   4.8\,\pm\, 1.0 $&$  0.25\,\pm\,  0.01 $&$  35.5\,\pm\,  0.6$ \\
$54.7$ & $   4.8\,\pm\, 1.1 $&$  0.22\,\pm\,  0.01 $&$  37.4\,\pm\,  0.5$ \\
$60.7$ & $   5.9\,\pm\, 1.0 $&$  0.20\,\pm\,  0.01 $&$  40.8\,\pm\,  0.5$ \\
$66.8$ & $   9.0\,\pm\, 1.5 $&$  0.17\,\pm\,  0.02 $&$  46.8\,\pm\,  0.5$ \\
\noalign{\smallskip}
\hline
\noalign{\smallskip}
$73.0$ & $  14.5\,\pm\, 0.5 $&$  0.65\,\pm\,  0.01 $&$  37.1\,\pm\,  0.6$ \\
$79.3$ & $   7.0\,\pm\, 1.0 $&$  0.72\,\pm\,  0.02 $&$  29.7\,\pm\,  0.7$ \\
$85.7$ & $  10.5\,\pm\, 1.1 $&$  0.74\,\pm\,  0.01 $&$  27.5\,\pm\,  0.8$ \\
$92.3$ & $  10.2\,\pm\, 1.3 $&$  0.66\,\pm\,  0.01 $&$  32.1\,\pm\,  0.9$ \\
$99.0$ & $  15.6\,\pm\, 1.1 $&$  0.57\,\pm\,  0.01 $&$  35.8\,\pm\,  0.9$ \\
\noalign{\smallskip}
\hline\hline
\noalign{\smallskip}
\end{tabular}
\end{table}

\subsection{Specific angular momentum $\lambda_R$}
\label{sec:lambda}
\begin{figure}
\resizebox{\hsize}{!}{\includegraphics{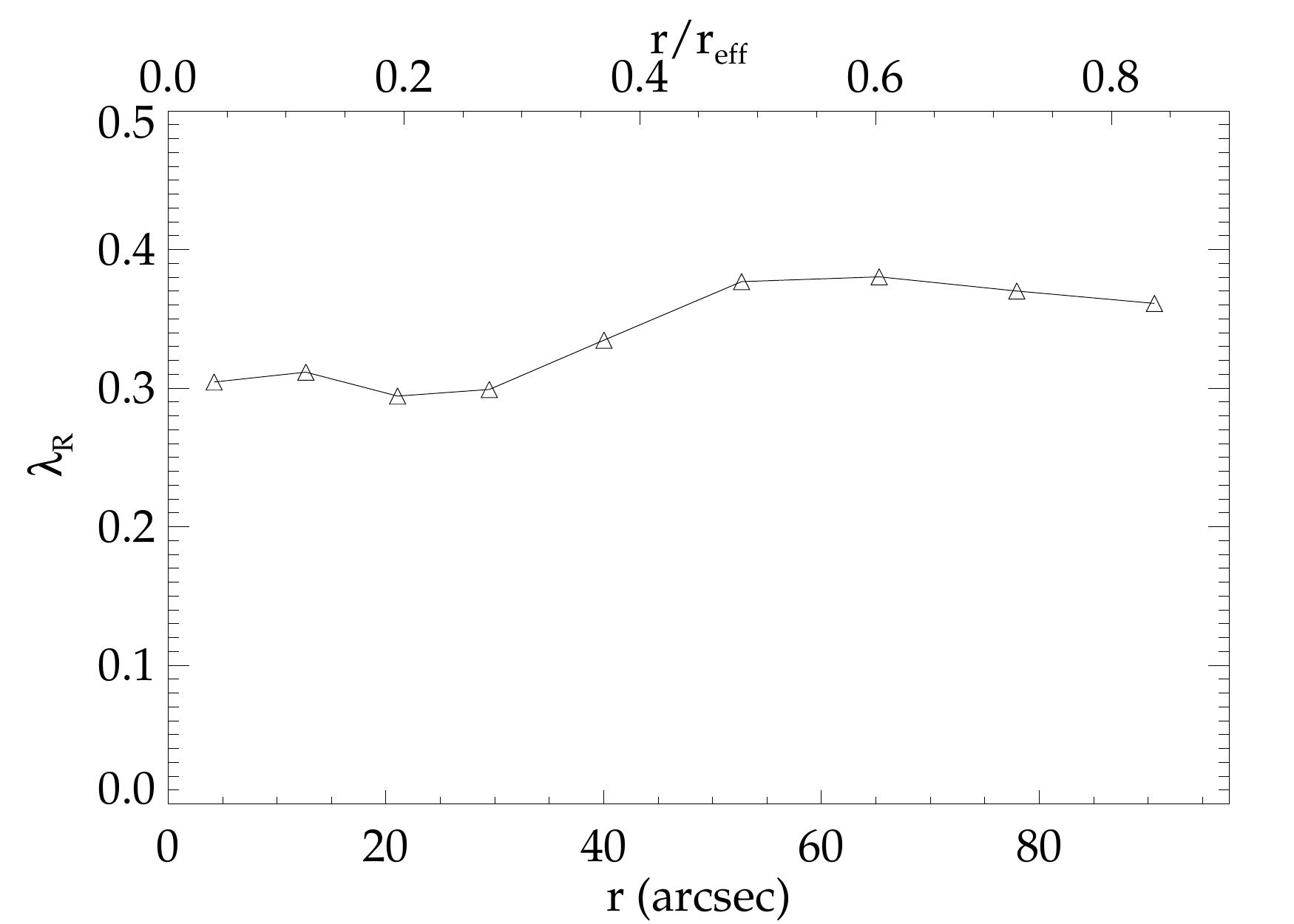}}
\caption{The specific angular momentum $\lambda_R$ profile calculated in ellipses with ellipticity $\epsilon$=0.29 \citep{sb}, and plotted against the mean radius of the ellipses. The upper x-axis of the plot displays the ratio of $r/r_{eff}$, where $r_{eff}$ is the effective radius \citep[$r_{eff}$=4.2\,pc,][]{sb}.} 
\label{fig:lambda}
\end{figure}

\cite{lambda} introduced the specific angular momentum $\lambda_R$ to classify early-type galaxies in slow and fast rotators. This is a more robust classification criterion than  the relation  $V/\sigma_e$. We calculate the  quantity $\lambda_R$ for the first time for a nuclear star cluster.  $\lambda_R$ is a dimensionless parameter that quantifies the observed projected stellar angular momentum per unit mass. To compute  $\lambda_R$ we use the two-dimensional spatial information from the kinematic maps, weighted by the luminosity. Slow rotators have values of $\lambda_R$\,\textless\,0.1 within their effective radius $r_{eff}$, while fast rotators have values above 0.1.  We calculate $\lambda_R$ using  Equation 6 of \cite{lambda}: 
\begin{equation}
\lambda_R = \frac{\sum_{i=1}^{N_p}\,F_i\,R_i\,\vert V_i \vert}{\sum_{i=1}^{N_p}\,F_i\,R_i\,\sqrt{V_i^2 + \sigma_i^2}}
\end{equation}
Here $i$ runs over all pixels of the data cube, $F_i$ is the flux in the $i-$th pixel, $R_i$ is its distance to Sgr~A*,  $V_i$ and $\sigma_i$ are the velocity and velocity dispersion in the pixel.
 To compute  $\lambda_R$ we use the SIRIUS image  and transform it to the same pixel size and astrometry as the data cubes for luminosity weighting. 
 We use the kinematic maps from the cleaned data cube, and sum within ellipses with ellipticity  $\epsilon$ = 0.29. This is the photometric ellipticity found by \cite{sb} in \emph{Spitzer} mid-infrared data.  
The resulting $\lambda_R$ profile is shown in Figure~\ref{fig:lambda}, where $r$ is the mean radius of the ellipses, and $r_{eff}$=4.2\,pc \citep[110\arcsec,][]{sb}.  Assuming a smaller value for $\epsilon$ decreases $\lambda_R$ and the slope of $\lambda_R$, but even with $\epsilon$  = 0, we obtain $\lambda_R$ \textgreater\,0.2  for all radii.

The specific angular momentum is used to classify galaxies as fast and slow rotators.  \cite{lambda2} found that fast rotators have $\lambda_R$\,\textgreater\,$k_{FS} \times \sqrt{\epsilon}$, where  $k_{FS}$ is a scaling parameter.   
  $k_{FS}$ is 0.31  when using $\lambda_{Re}$, i.e.  $\lambda_R$ at one effective radius $r_{eff}$. This means a value of $\lambda_{Re}$\,\textgreater\,0.16 would be expected for a fast rotator with an  ellipticity of 0.29. For the Milky Way nuclear star cluster $r_{eff}$ is  $\sim$4.2\,pc  \citep{sb}. At our outermost ellipse radius of 3.5\,pc we obtain $\lambda_R \approx $ 0.36 with $\epsilon$ = 0.29.  So the Milky Way nuclear star cluster has similar rotational support as fast rotating elliptical galaxies have. A comparison with Figure~5 of \cite{lambda} shows that $\lambda_{Re}$  is below the value expected for an isotropic oblate rotator. This suggests that the Milky Way nuclear star cluster has anisotropic kinematics at large radii. 

\subsection{Radial  profiles} 
\label{sec:sig}
In this section we study the kinematics of the nuclear star cluster by applying different binnings to the data cubes and compute profiles. 
We investigate the rotation curve and compare our findings to previous results.

\begin{figure*} 
\centering
\includegraphics[width=18cm]{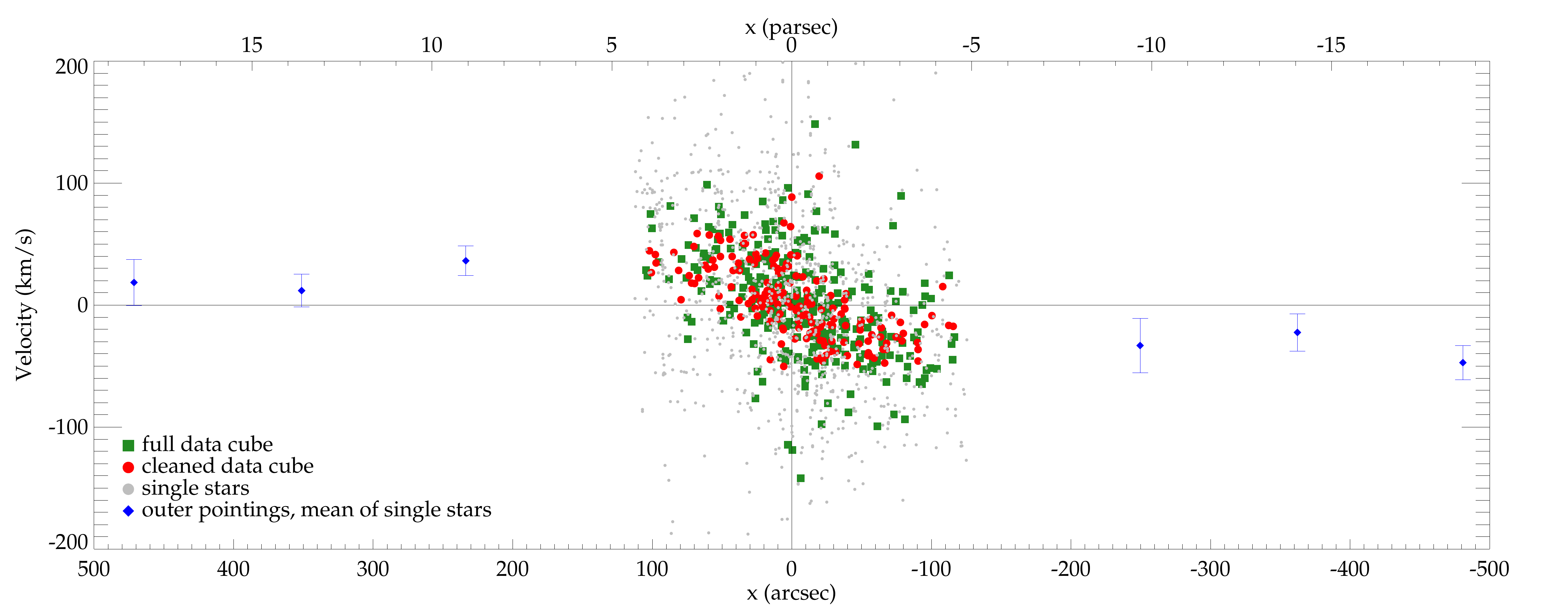}
\caption{Position-velocity diagram along the Galactic plane for the velocity  maps of Figure~\ref{fig:map60} and \ref{fig:map_mag60}.  Results from  our full ISAAC data cube  are shown as green rectangles,  red circles show the result from the cleaned data cube, which  contains only faint  cluster member stars. Grey dots denote velocity measurements from single stars, the blue diamonds are the mean velocities of the bright stars at the outer fields, corrected for perspective rotation. }
\label{fig:posvel}
\end{figure*}
When we  plot the projected distance of each Voronoi bin from Sgr~A* along the Galactic plane against the velocity of each bin, we obtain a position-velocity diagram, which is shown in Figure~\ref{fig:posvel} for both data cubes.  
From the data cube containing only faint member stars (red circle points), we deduce that the maximum rotation velocity is about 50\,km/s. For the outer fields, the rotation is flatter. Only one of the six data points has a velocity higher than 36\,km/s.  

  \begin{figure}
\resizebox{\hsize}{!}{\includegraphics{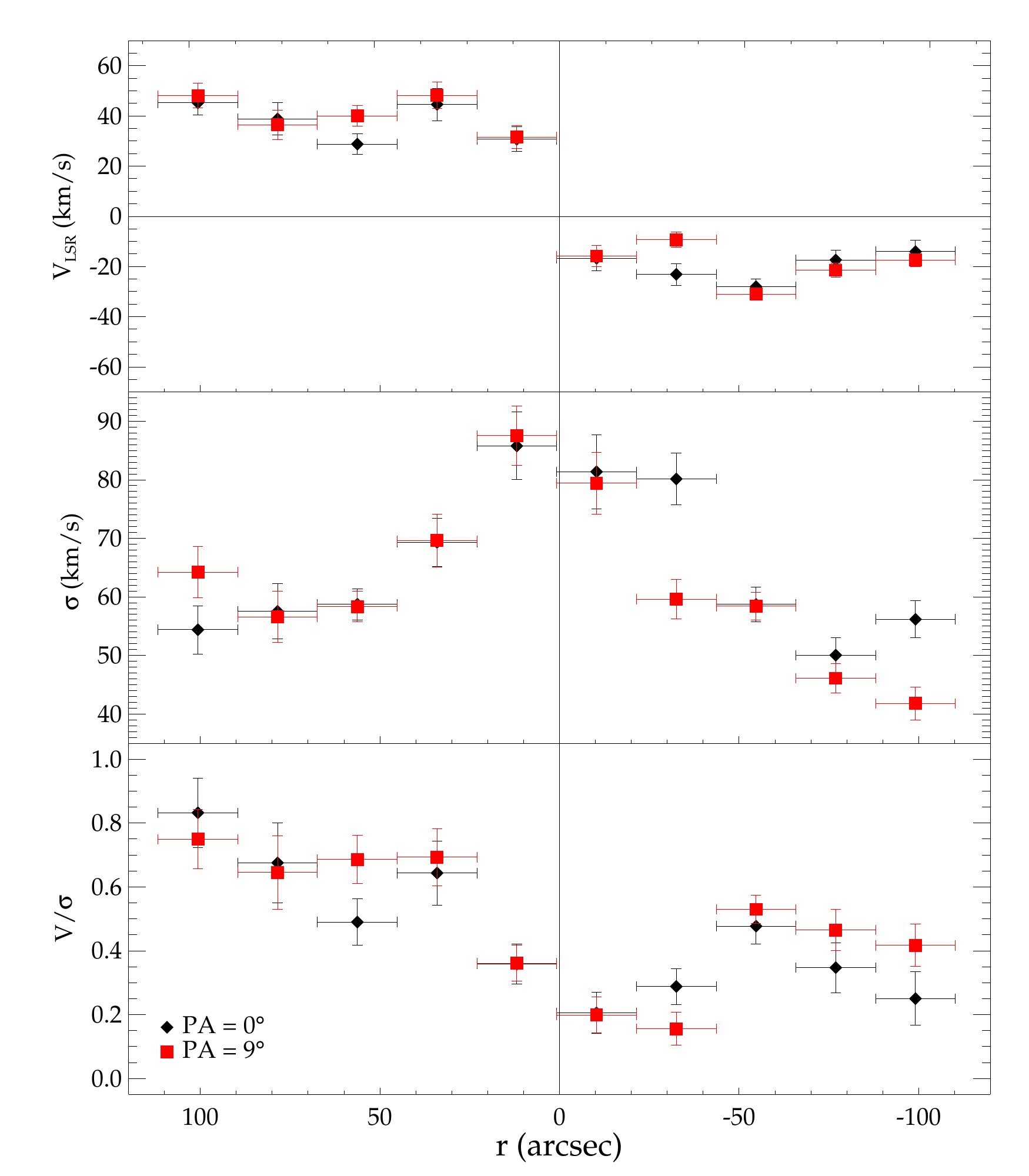}}
\caption{Velocity profile, velocity dispersion $\sigma$, and $V/\sigma$ for a 22\arcsec\,broad slit along the Galactic plane from the cleaned data cube (black diamonds), and for a slit tilted by 9\degr\,counterclockwise with respect to the Galactic plane (red square symbols). }
\label{fig:3prof}
\end{figure}

For further analysis we do not use the Voronoi binned velocity maps of Figure~\ref{fig:map60} and \ref{fig:map_mag60}, but directly bin the data cubes in major axis bins and radial annuli.
 In order to derive the kinematics along the photometric major axis only, we bin the cleaned data cube in rectangles, which extend  over $\sim$22\arcsec~perpendicular to the Galactic plane, i.e. $\sim$11\arcsec~towards the Galactic North and South of Sgr~A*. Each rectangle is   22\arcsec~broad in Galactic east-west direction. This resembles a slit along the Galactic plane, centred on Sgr~A* that is 22\arcsec~wide. We sum the spectra in the rectangular bins and fit the CO absorption lines with \emph{ppxf}. The resulting velocity and velocity dispersion profiles are shown in Figure~\ref{fig:3prof} as black diamonds, together with the $V/\sigma$ profile. Alternatively we  rotate the slit axis by  9\degr, which is our median kinematic position angle offset found in Section~\ref{sec:intlight}, and apply a similar binning. The result is  shown as red square symbols in Figure~\ref{fig:3prof}.

We obtain a  rotation curve with an amplitude of $\sim$40\,km/s. As expected, the  absolute velocity values   are generally  higher using the tilted slit than using a slit along the Galactic plane.  Further, the absolute velocities are higher on the eastern side than on the western side by more than 10\,km/s. 
The velocity dispersion rises only to  $\sim$\,\!90\,km/s. This can be explained by the fact that the velocity dispersion map (Figure~\ref{fig:map_mag60}) has values higher than 100\,km/s only in few small bins in the very centre (r $\lesssim$ 10\arcsec).  The V/$\sigma$ has a minimum close to the centre, where the velocity dispersion is highest, and higher values at larger radii. At  outer  radii (r \textgreater 50\arcsec $\approx$ 2\,pc), closer to the effective radius,   we obtain V/$\sigma \approx$ 0.6. The value of V/$\sigma$ at the Galactic West is smaller than at the Galactic East, as also   the absolute values of the velocities are lower.

\begin{table}
\caption{Profile of  the root-mean-square velocity  $V_{rms}$ as shown in Figure~\ref{fig:dprof}. $r$ is the distance from Sgr~A* in arcsec, $V_{rms}$ values are in km/s. }
\label{tab:dprof}
\centering
\begin{tabular}{c c c }
\noalign{\smallskip}
\hline\hline
\noalign{\smallskip}
$r$&$ V_{rms} $&$V_{rms} $\\
$\left[ arcsec \right] $ & $\left[ km/s \right] $ &$\left[ km/s \right] $  \\
&full data cube& cleaned data cube\\

\noalign{\smallskip}
\hline
\noalign{\smallskip}
$1.5$ & $  179\,\pm\, 20 $&$  166\,\pm\,  20 $\\
$4.0$ & $   14\,\pm\, 2$&$ 136\,\pm\,  21 $\\
$7.5$ & $   105\,\pm\, 8 $&$  98\,\pm\,  10 $ \\
$12.5$ & $  78\,\pm\, 4 $&$  73\,\pm\,  5 $\\
$20.0$ & $  78\,\pm\, 3 $&$  84\,\pm\,  4 $ \\
$32.5$ & $   78\,\pm\, 4$&$  71\,\pm\,  3 $ \\
$47.5$ & $  67\,\pm\, 3 $&$  72\,\pm\,  2 $ \\
$62.5$ & $   65\,\pm\, 2$&$  67\,\pm\,  3 $ \\
$77.5$ & $  65\,\pm\, 2 $&$  62\,\pm\,  3 $ \\
$92.5$ & $   69\,\pm\, 2$&$  67\,\pm\,  3 $ \\
$107.5$ & $  69\,\pm\, 2 $&$  61\,\pm\,  3 $ \\
$122.5$ & $  71\,\pm\, 3 $&$  69\,\pm\,  5 $ \\
\noalign{\smallskip}
\noalign{\smallskip}
$255$ & $  75\,\pm\, 3 $&$  58\,\pm\, 4 $ \\
$355$ & $   49\,\pm\, 4 $&$  77\,\pm\,  5 $ \\
$455$ & $  78\,\pm\, 3 $&$  74\,\pm\,  4 $ \\
\hline\hline
\noalign{\smallskip}
&single stars& \\
$241$ & $  87\,\pm\, 9 $&$   			$ \\
$357$ & $  67\,\pm\, 7 $&$  $ \\
$455$ & $  79\,\pm\, 9 $&$    $ \\
\noalign{\smallskip}
\hline\hline
\noalign{\smallskip}
\end{tabular}
\end{table}

\begin{figure}
\resizebox{\hsize}{!}{\includegraphics{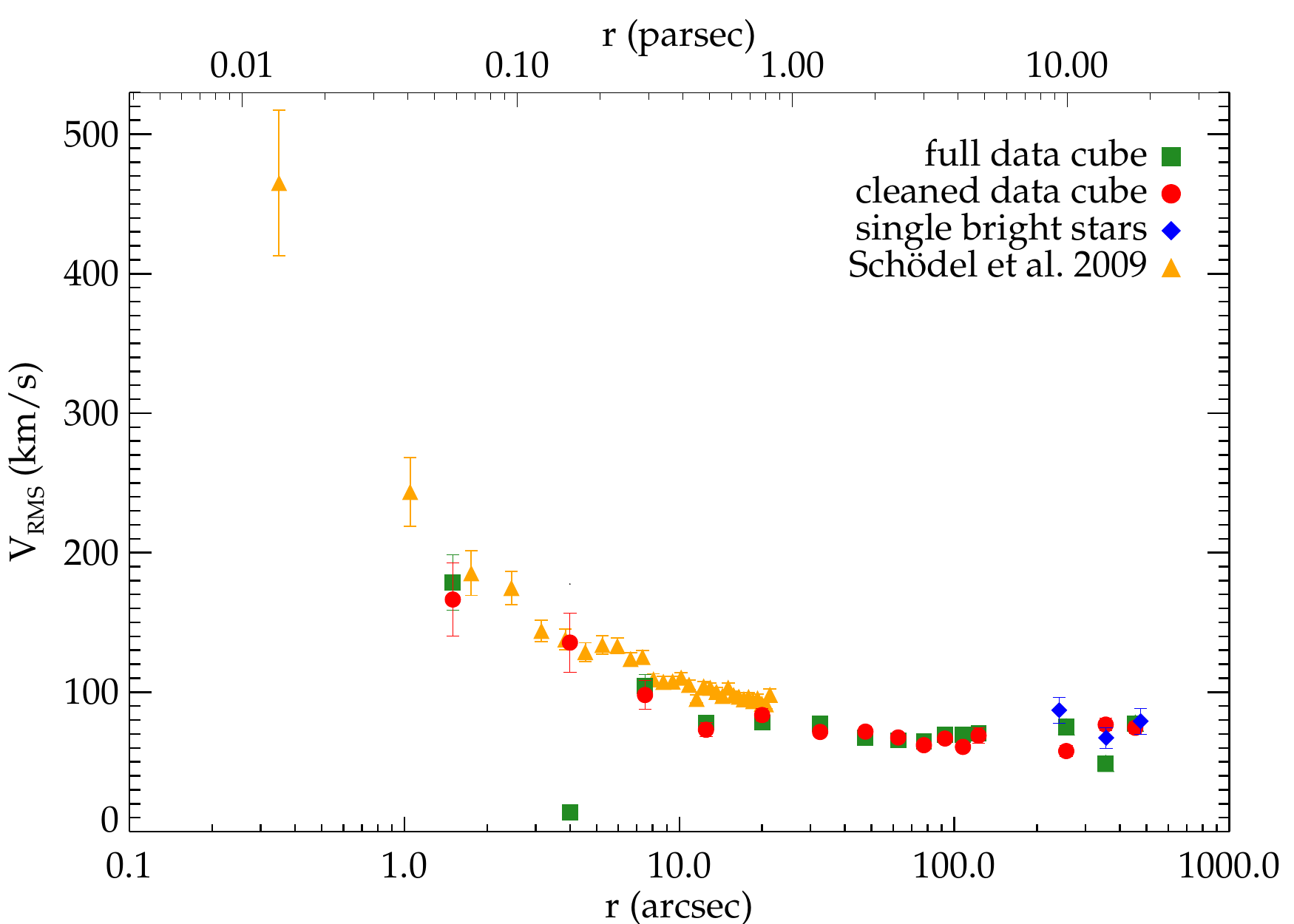}}
\caption{Velocity  dispersion profile, obtained using circular bins around Sgr~A*.  The green rectangles are the data points from the full data cube, red circle points are from the cleaned data cube.   The large discrepancy at $r$ $\sim$\,4\arcsec\,is due to the influence of the star IRS~7 on the green point. Blue diamonds are obtained from single bright stars in the outer fields. Orange triangles are proper motion data from \cite{Rainerpm09}.}
\label{fig:dprof}
\end{figure}
We also compute a velocity dispersion profile within circular bins around Sgr~A*. The spectra in each bin are summed up and we compute the LOSVD with \emph{ppxf}. The second moment is then the root-mean-square velocitiy,  $V_{rms} = \sqrt{V^{2}+\sigma^{2}}$. The result is shown in Figure~\ref{fig:dprof} and listed in Table~\ref{tab:dprof}, the outermost data points are from our outer fields.  
Green rectangles denote the result from the full data cube, the red circle points from the cleaned data cube. 
The difference between the two data cubes is mostly within the measurement uncertainties.  The largest differences occur due to contamination of IRS~7 in the bin at 4\arcsec\,and another bright star at a radius of 360\arcsec.  
Most stars in the centre of the nuclear star cluster are young and do not show strong CO absorption lines that we use for our kinematic measurements. This means that the velocity dispersion we measure for the central points comes mostly from stars that lie only in projection close to Sgr~A*, and the velocity dispersion at $r$\,$\lesssim$\,8\arcsec\,is  a lower limit \citep{sellgren90,haller96}.

The surface brightness at the outer fields is lower than in the central field. After cleaning the data from bright stars and foreground stars, the  S/N of the outer fields is rather low. On the other hand, bright stars can severely effect the result from the full data in the outer fields. Therefore we compute $V_{rms}$ as a third approach from the bright stars we cut out, corrected for perspective rotation (see Section~\ref{sec:co}). This has the advantage that all stars have the same weight, irrespective of their magnitude. The results from the three approaches agree quite well at 8\arcmin; at 4\arcmin\,and 6\arcmin\, there is more scattering.

For the central 20\arcsec\,we also plot the proper motion data of \cite{Rainerpm09} for comparison (orange triangles). To extract these data along a radial profile we make circular bins around Sgr~A*. Then we compute the mean velocity and velocity dispersion of each bin using a maximum likelihood approach as described in \cite{pryor}. In contrast to computing the  mean and standard deviation of the data, this method takes the different uncertainties for the velocity measurements into account. We use the velocities perpendicular and parallel to the Galactic plane. Each bin contains at least 21 stars. We found that the velocity dispersion perpendicular to the Galactic plane $\sigma_b$ is in better agreement with our LOSVD data than the velocity dispersion parallel to the Galactic plane $\sigma_l$. Parallel to the Galactic plane the velocity dispersion is higher than perpendicular to it \citep[see Figure~6 of ][]{Rainerpm09}.

\begin{figure}
\resizebox{\hsize}{!}{\includegraphics{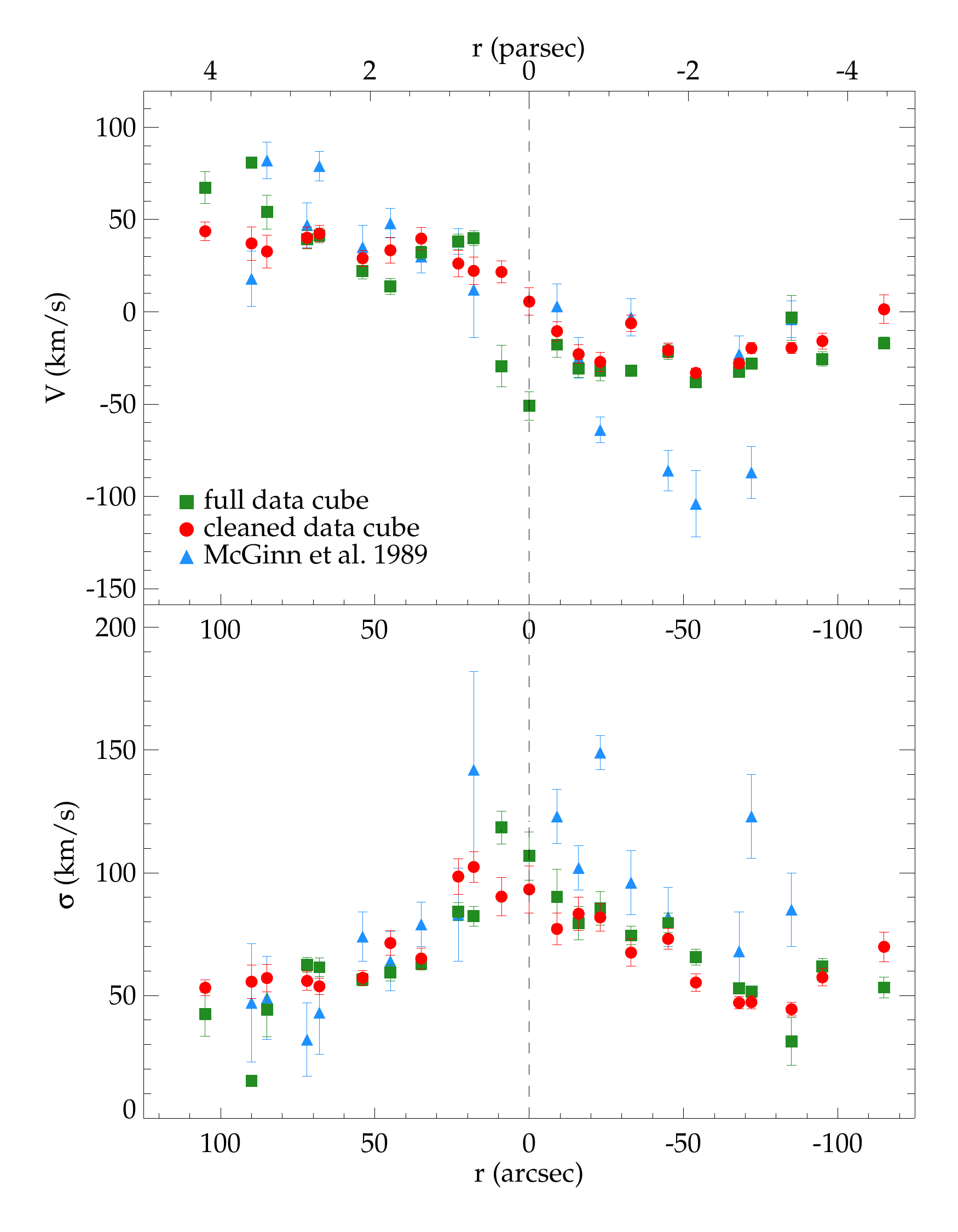}}
\caption{Velocity (upper panel) and velocity dispersion profile (lower panel) along the major axis, with IRS 16 as the centre. We  applied the binning of \citealt{mcginn89} (blue), extended by five additional binning apertures.  Results from  our full ISAAC data cube  are shown as green rectangles,  red circles show the result from the cleaned data cube, and blue triangles denote the data from \cite{mcginn89}.}
\label{fig:mcginn}
\end{figure}

The only previous observations of  the Galactic centre in integrated light  are from \cite{mcginn89}. They had a circular aperture of 20\arcsec\,diameter, and  measured velocity and velocity dispersion profiles along the major axis \citep[Figures 1, 4, and 5]{mcginn89} using the first CO band head. To compare our work with these early results, we bin our data cube in exactly the same way as \cite{mcginn89}. As centre we use IRS 16C \citep[$\alpha$=266.41705\degr, $\delta$= -29.007826\degr,][]{irs16}, as \cite{mcginn89} did.  We apply four additional binnings which are not covered by \cite{mcginn89}. The resulting profiles are displayed in Figure~\ref{fig:mcginn}, where the upper panel is the velocity V$_{LSR}$, and the lower panel is the velocity dispersion $\sigma$, both in km/s. 
  The  profiles from the cleaned data cube appear smoother than the profiles from the full data cube, as the data is less affected by shot noise of dominant bright stars which can produce outliers. One such example is the bright supergiant IRS~7. 
  This star  has a big effect on the integrated spectrum in the bins centred  at $r = -$9\arcsec\,and 0\arcsec. Cutting out only the star IRS~7 from the data cube containing all stars decreases the velocity measurement by more than 30\,km/s,  while the velocity dispersion changes by only $\sim$\,14\,km/s. Using smaller apertures, this single bright star can affect the results even more.
The data of \cite{mcginn89} has more outliers and extreme velocity variations than our data, especially to the western side of Sgr~A*. \cite{Rainerpm09} suggested that the data of \cite{mcginn89} may sample two different stellar populations at different depth. Then their data could hint at two distinct rotating systems in the nuclear star cluster. 
Our data does not confirm this large scatter in the velocity and velocity dispersion,  and the profiles have a rather smooth shape. Four velocity measurements of \cite{mcginn89} on the western side of Sgr~A* have significantly higher absolute values than our data has, while the other five measurements agree with our data. They confirm that the absolute values of the velocities on the western side are lower than on the eastern side of Sgr~A*. On the western side the extinction  is higher than on the eastern side.

\subsection{Distribution of young stars}
\label{sec:young}
\begin{figure*}
\centering
\includegraphics[width=18cm]{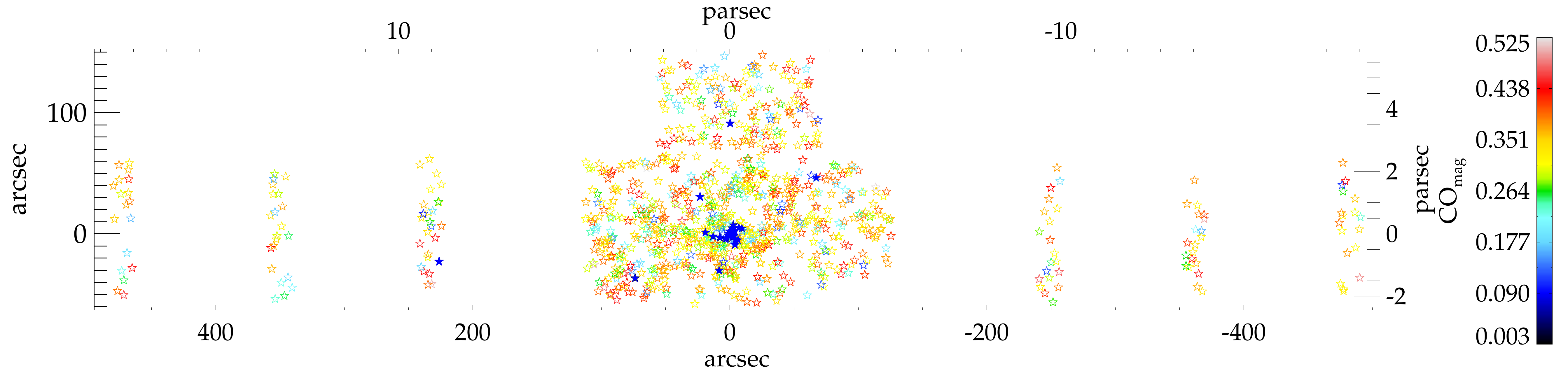}
\caption{CO index map of  1,153 bright stars (K $\le$ 11.5$^m$) with S/N \textgreater 12. Filled blue star symbols denote stars with $CO_{mag}$ \textless\,0.09, which are younger than 300\,Myr.  }
\label{fig:map_co}
\end{figure*}

 We compute the CO index $CO_{mag}$ for our sample of bright stars as defined in  Section~\ref{sec:coindex}. Figure~\ref{fig:map_co} shows the distribution of the CO index for all bright member stars. The value of $CO_{mag}$ roughly correlates with the age of the stars, this means that blue data points denote stars that are more likely to be younger than stars shown in red colours.  Blue filled star symbols denote stars with $CO_{mag}$\,\textless\,0.09, indicating stars younger than 300\,Myr. We note that the presence of foreground extinction would mean the stars we are sampling are even brighter and thus younger. Most of the 26 young stars are within 1\,pc of the nuclear star cluster, confirming previous studies. For example, five of our stars lie within 0\farcs3 distance of stars listed by \cite{paumard06}. 
A comparison with the young star candidates of \cite{shogo12} results in seven matches. One of those matches has no counterpart in any previous spectroscopic study yet.  Its colour in H$-$K=1.5$^m$, which is at our limit for foreground stars. 
At greater distance from Sgr~A*, there are only few stars with $CO_{mag}$\,\textless\,0.09, 
and none at our outer fields at $\sim$\,\!360\arcsec\,and $\sim$\,\!480\arcsec. But as the covered area beyond 350\arcsec\,is only $\sim\,\!2.1\,arcmin^2$, we might simply miss the young stars.  

It is possible that the young star candidates are foreground stars, although their colours in H$-$K are at least 1.5$^m$, and for 20 of the  stars even more than 2.1$^m$.  The intrinsic H$-$K colour for  O to A type stars  is about $-0.1$ to 0.05 \citep{intrinsiccolor}, therefore the measured  colour is almost entirely caused by extinction. \cite{distance} and \cite{fritzextinktion} found mean extinction values of $A_H$=4.21$^{m}$ to 4.48$^{m}$ and $A_{Ks}$=2.42$^{m}$ to 2.74$^{m}$ in the central region of the Galaxy. Extinction varies on arcsecond scale with a standard deviation of  0.3$^{m}$ in $A_{Ks}$ \citep{distance}. Thus the measured colours are consistent with these stars being located in the Galactic center.

One problem of our limited spectral range is that stars with velocities of less than $-$250 km/s can be misidentified as a star with a low CO index. We found three such stars, which are not included in our sample of young star candidates. Observations with  larger spectral range are necessary to confirm the spectral type of these stars and give a better estimation of their age.

\section{Dynamical modelling}
 \label{sec:sbjeans}
To measure the mass of the Milky Way nuclear star cluster and black hole we dynamically model our kinematic data.
 We apply axisymmetric Jeans models \citep{jeans} to our kinematic data. 
  We use the IDL program package  \textit{Jeans Anisotropic MGE dynamical models} (JAM) written by \cite{jam}. In Section~\ref{sec:sb} we fit a surface brightness profile. We present the results of the  Jeans models in Sections \ref{sec:2dmodel}. 
 Our mass profile is shown in Section~\ref{sec:massprofile1}.

\subsection{Surface brightness profile}
\label{sec:sb}
Light indicates the stellar density, and for a constant mass-to-light ratio (M/L) also the  gravitational potential of a stellar system. For the Jeans models we require a  surface brightness profile.  We use mid-infrared data to determine the surface brightness profile, as the interstellar extinction towards the Galactic centre reaches a minimum at $\sim$\:\!3$-$8$\,\mu m$\, \citep{fritzextinktion}.  We derive a surface brightness profile using imaging data from Spitzer/IRAC \citep{iracim} at 4.5\,$\mu m$. The image is dust and extinction corrected, and smoothed to a  pixel scale of 5\arcsec/pixel. For details about the correction steps we refer to \cite{sb}.
For the centre, we use a NACO $K$-band mosaic \citep{Rainerpm09} covering 40\farcs5~$\times$~40\farcs5 scaled to the flux of the 4.5\,$\mu m$ \emph{Spitzer} image, with a pixel scale of 0\farcs027/pixel. 

We use the \emph{$MGE\_\,FIT\_\,SECTORS$} package. This is a set of IDL routines written by \cite{mgeidl} to do photometric measurements directly on images and apply a Multi-Gaussian Expansion \citep[MGE,][]{mgeeric} fit  to parameterise a surface brightness profile.  The photometry of the two images is measured, and we determine a scale factor using the overlap region of the two images to convert the NACO image to the \emph{Spitzer} flux. Then photometry is measured on each image along 12 angular  sectors. 
The routine assumes four-fold symmetry and averages the measurements of four quadrants, taken along elliptical annuli with constant  ellipticity $\epsilon$.  

A set of two-dimensional Gaussian functions is fitted to the combined photometric measurements. Therefore we also take the PSF of the NACO image into account. The MGE fit is not designed for a structural decomposition of the Galaxy's light profile. But 
Gaussians have the great advantage that deprojection can be done analytically. This procedure has been used already for  galaxies \citep[e.g.][]{mgeanwendspher}, nuclear clusters \citep[e.g.][]{anil10404}, and globular clusters \citep[e.g.][]{mgeanwendglob,5286anja}.

\begin{figure}
\resizebox{\hsize}{!}{\includegraphics{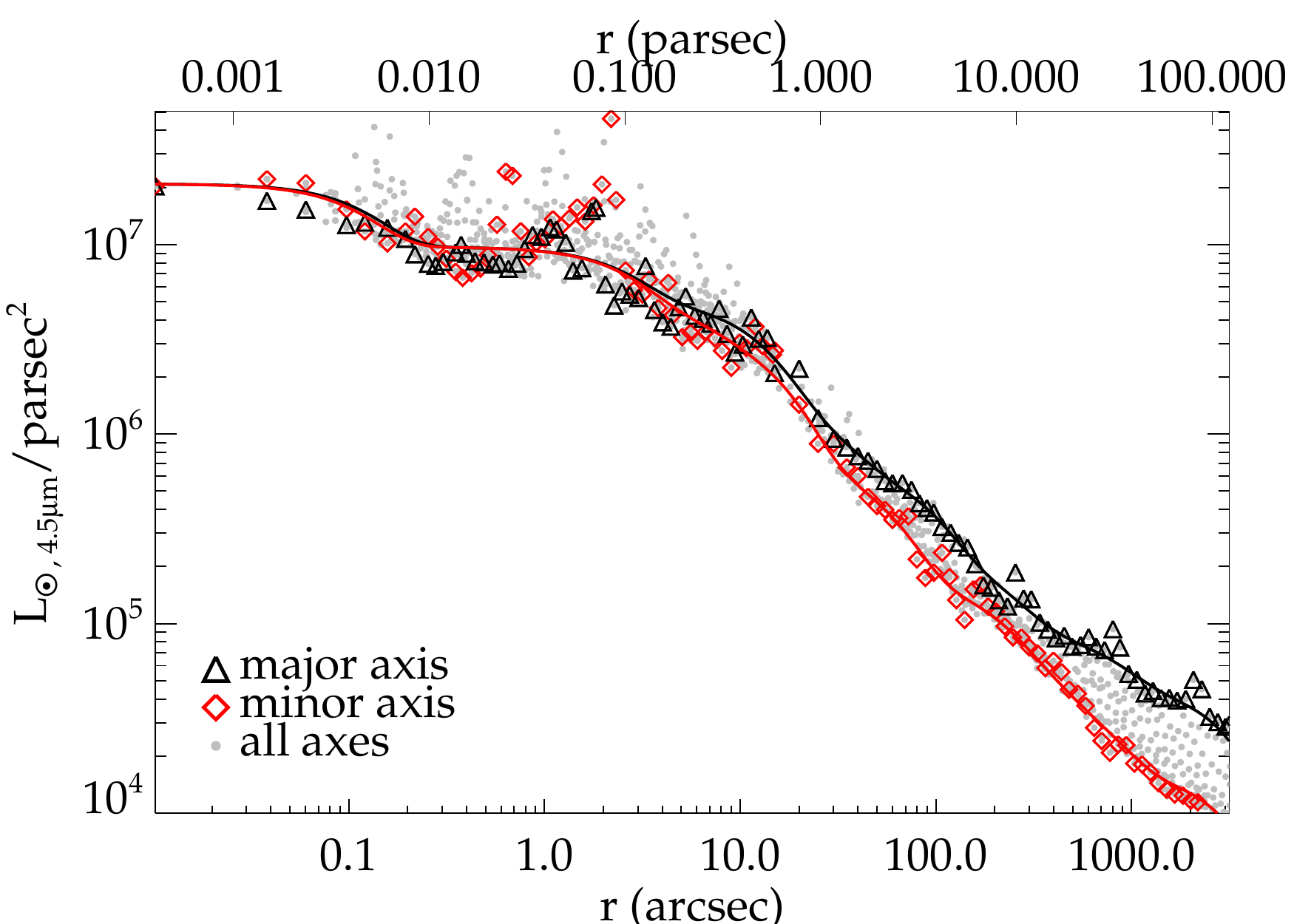}}
\caption{Surface brightness profile derived from a dust and extinction corrected Spitzer/IRAC 4.5\,$\mu m$ image and NACO $K$-band mosaic for the central 30\arcsec. Black triangles denote measurements along the major axis, and red diamonds along the minor axis, the grey data points are all measurements, at all axes. The spikes in the profile are due to individual bright stars. Solid lines illustrate the Multi-Gaussian Expansion (MGE) fit to the data.  }
\label{fig:sb}
\end{figure}
\begin{table}
\caption{The Multi-Gaussian Expansion (MGE) fit parameters for the 4.5$\mu m$ Spitzer/IRAC dust and extinction corrected image in combination with the  NACO $K$-band mosaic scaled to \emph{Spitzer} flux. $I$ is the peak surface brightness, $\sigma_{MGE}$ is the standard deviation, and $q$ is the axial ratio of the Gaussian components.}
\label{tab:sbprof}
\centering
\begin{tabular}{c c c }
\noalign{\smallskip}
\hline\hline
\noalign{\smallskip}
$I $ &$\sigma_{MGE}$  &$ q$\\
$\left[ 10^5 L_{\odot,4.5\mu m} / pc^2 \right] $  & $\left[ arcsec \right] $   &  \\
\noalign{\smallskip}
\hline
\noalign{\smallskip}
$   112	$&$	 0.1$&$   0.9 $\\
  $     46.2   $&$    2.1   $&$ 1.0$\\
  $     16.0    $&$   8.4    $&$    0.6$\\
   $    20.4    $&$   11.5     $&$  1.0$\\
    $   7.48     $&$  22.8     $&$ 0.7$\\
   $    4.53   $&$    66.4     $&$  0.7$\\
     $  0.77    $&$   143     $&$  1.0$\\
   $    0.73    $&$   184      $&$  0.2$\\
     $  0.47    $&$   581     $&$  0.4$\\
   $    0.17    $&$   2705      $&$  1.0$\\
 $    0.31    $&$   2705      $&$  0.2$\\

\noalign{\smallskip}
\hline
\noalign{\smallskip}
\end{tabular}
\end{table}
We assume the position of Sgr~A* as centre and exclude the dark 20\,km/s cloud  and the Quintuplet cluster from the \emph{Spitzer} image. As the central region of the 4.5\,$\mu m$\, emission is dominated by the mini-spiral and not by stellar emission, we also exclude pixels within  0.6\,pc distance to Sgr~A* of the \emph{Spitzer} image from photometry measurements. The flux scaled NACO image is used out to 15\arcsec\,distance from Sgr~A*. 
Table~\ref{tab:sbprof} lists the output of the MGE fit.  The last two Gaussian components have the same value of $\sigma_{MGE}$, as they are close to the edge of the image.  Figure~\ref{fig:sb} shows  the MGE surface brightness profile along the  major and minor axes. 

 \subsection{Axisymmetric Jeans models}
 \label{sec:2dmodel}

 For the  axisymmetric Jeans models we use the surface brightness parametrisation of  Table~\ref{tab:sbprof} as an input. We assume an inclination angle of 90\degr, i. e. the Galaxy is seen edge-on. We fit the model to the kinematic data. For this purpose we  use the velocity maps of Figure~\ref{fig:map_mag60} of the faint stellar population, and exclude the bin with the highest uncertainty and lowest S/N. From these maps we compute the root-mean-square velocity  $V_{rms} = \sqrt{V^{2}+\sigma^{2}}$ for each bin. In all models we assume  a constant mass-to-light ratio  M/L.  All M/L values are for 4.5\,$\mu m\,$ (M/L$_{4.5 \mu m}$) and in units of $M_\odot/L_{\odot,4.5 \mu m}$.
 
 For the anisotropy $\beta$ (= 1$-\overline{v_z^2}/\overline{v_R^2}$, \citealt{jam}) we test different assumptions. We assume isotropy in the central 0.5\,pc \citep{tuan}, but fit the anisotropy further out. Therefore we assume different radial shapes of the anisotropy $\beta$: 1) constant anisotropy beyond 0.5\,pc; 2)  logarithmically increasing anisotropy beyond 0.5\,pc, i.e. $\beta \propto \log(r+0.5\,pc)$; 3) linear increasing anisotropy beyond 0.5\,pc, i.e. $\beta \propto r$.   We have three fit parameters:  M/L, $\beta_{100pc}$, and the black hole mass M$_{\bullet}$. As a fourth approach, we assume constant anisotropy over the entire cluster and do not constrain the central 0.5\,pc to be isotropic and fit M/L, $\beta_{const}$, and the black hole mass M$_{\bullet}$. We limit the value  of $\beta$ to [-4; 0.9], since lower and higher values seem to be unrealistic.

\begin{figure*}
\centering
\includegraphics[width=18cm]{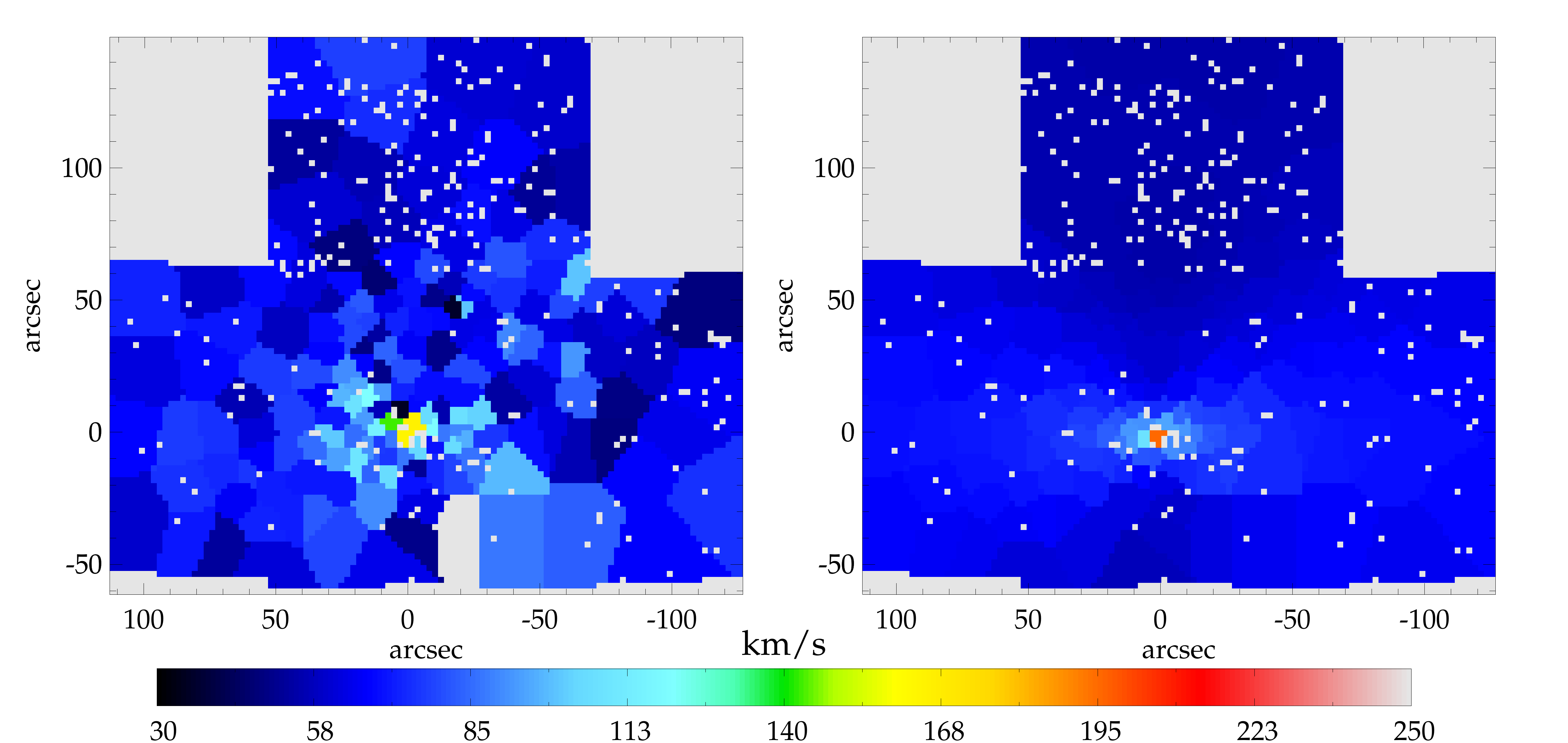}
\caption{Result of the two-dimensional Jeans model with anisotropy $\beta_{const}$ =  $-$0.3,  mass-to-light ratio M/L  = 0.56, and black hole mass M$_{\bullet} = 1.7\, \times\,10^6$\,M$_\odot$. The left panel shows ISAAC data  of the  cleaned cube,  for the root-mean-square velocity  $V_{rms} = \sqrt{V^{2}+\sigma^{2}}$,  and the right panel is the best-fit model. The grey pixels mark regions, which contain no information. One bin was excluded from the fit due to its high uncertainty and low signal-to-noise, and it is also shown in grey on the left panel.}
\label{fig:2din}
\end{figure*}

  Our best fit model has constant tangential anisotropy, the best fit parameters are $\beta_{const}\,\!=\,\!-0.3 ^{+0.3}_{-0.4}$,  M/L $\,\!=\,\!0.56 ^{+0.22}_{-0.26}$, and M$_{\bullet}\,\!=\,\!(1.7 ^{+1.4}_{-1.1})\,\!\times\,\!10^6$\,M$_\odot$, with $\chi_{red}^2$ =  12.5.  The uncertainties are the  68\% confidence limits. Data and model are shown in Figure~\ref{fig:2din}. The left panel is the data, the right panel displays the best-fit model.    
 
 The results quoted here are from models without PSF convolution as the effect is negligible in our low spatial resolution data. In order to test this  assumption we performed a PSF convolution with the seeing of the ISAAC data and found a difference that was less than 10\% of the uncertainties.  A variation of the inclination angle to 80\degr\, has no effect on the best fit results, but increases  $\chi_{red}^2$ slightly.

The different $\beta$-profiles obtain consistent results for M/L and M$_\bullet$, with only slightly higher values of $\chi_{red}^2$  (13.0$-$13.4). This suggests that we cannot constrain a possible $\beta$ variation over the cluster radius with our  Jeans models. 
There is not much  difference in the results if  we run our models with the six outer fields or without. Using the kinematic maps containing all stars (Figure~\ref{fig:map60}), the tangential anisotropy is increased ($\beta_{100pc}$ and $\beta_{const}$\,\!$\le$\,\!$-$0.5),  the best fit M/L is lower ($\sim$0.32), and the black hole mass is slightly higher ($\sim$2.2\,$\times~10^6$\,M$_\odot$). Differences to the results from the kinematic map of the cleaned data cube are less than the uncertainty limits. The $\chi_{red}^2$ is much higher with values between   235 and 242, since the kinematic map of the full data cube is more affected by shot noise caused by single bright stars. 

As the population in the central parsec  (r $\sim$13\arcsec) is younger than at larger distances from Sgr~A*, this might influence the result of the M/L.  We tested this by running the Jeans model without the $V_{rms}$ data points inside $r$ = 10\arcsec\,and $r$ = 20\arcsec. In both cases there is no difference in the best fit values of M/L, M$_\bullet$ and $\beta$.  As another test we set the M/L of the innermost five Gaussian components ($\sigma_{MGE}$ \textless 0.9\,pc) to zero before we run the Jeans models. This results in an increase of the M/L of the outer components to 0.7  and M$_\bullet$ to 3.4$\times 10^6$\,M$_\odot$. The value of M/L is within the uncertainty limits, while the black hole mass fit is significantly improved. But even with such an extreme assumption of M/L=0 in the central parsec, the black hole mass is too low. 
 When we force the  black hole mass to be 4\,\!$\times$\,\!10$^6$\,M$_\odot$, the best-fit value of M/L is 0.35 and $\beta$ = $-$0.3. This is within our uncertainties for M/L.

\subsection{Mass profile}
\label{sec:massprofile1}
 
With the mass-to-light ratio and our surface brightness profile we  compute the enclosed mass of the nuclear star cluster.  The light profile is parameterised by a series of Gaussians, which is given in Table~\ref{tab:sbprof}. This profile is valid out to 100\,pc distance from Sgr~A*. We compute the enclosed light within ellipses for each Gaussian component with the equation 
\begin{equation}
L_{j}  =  2\, \pi\, I_j\, \sigma_{MGE,j}^2\, q_j\, \left( 1-\exp{\left( -\frac{x_{max}^2}{2\sigma_{MGE,j}^2}\right)} \right),
\end{equation}
where $x_{max}$ is the semi-major axis distance from the centre, up to which we integrate the light. Figure~\ref{fig:encmass} illustrates the enclosed mass as a function of the mean radius of an ellipse $r=x_{max}\times~\sqrt{1-\epsilon}$ with $\epsilon$=0.29   for a M/L of 0.56$^{+0.22}_{-0.26}$. This is the best fit result of Section~\ref{sec:2dmodel}. 
The uncertainty of our light profile, which we get from the uncertainty map of the \emph{Spitzer} image, has about the same effect on the enclosed mass profile as a M/L variation of 0.1.

At a distance of 10\,pc from Sgr~A*, the nuclear star cluster is still the dominant component of the Galaxy \citep{launhardt02}, it contributes more than the nuclear stellar disk. We obtain an enclosed mass of (3.0$^{+1.2}_{-1.4}$)\,$\times$~10$^7$~M${_\odot}$  at a mean  distance of 10\,pc. 
While this mass is an extrapolation from the area where we have good sampling of the kinematics, it is more robust than previous dynamical estimates \citep{mcginn89,lindqvist922,trippe08,Rainerpm09}.

\section{Discussion}

\label{sec:summary}
 \subsection{Clues to the formation of the Milky Way nuclear star cluster}
 \label{sec:disussionkin}

Our velocity maps of the Milky Way nuclear star cluster reveal two unexpected features: (1) the offset of the rotational position angle from the Galactic plane and photometric position angle of the nuclear star cluster, and (2) indications for the presence of a rotating substructure at  $r$ $\sim$ 0.8\,pc rotating perpendicular to the Galactic plane.

Our kinemetric analysis  of the velocity map in Section~\ref{sec:intlight} revealed an offset between the photometric and the kinematic position angle with a median value of  9\degr$\pm$3\degr\,Galactic east of north. The photometric position angle is at $\sim$0\degr\,\citep{sb}.  The misalignment between kinematics and morphology suggests that the overall luminosity profile is dominated by a different stellar population than the kinematics. 
The exact value of the position angle offset depends on the binning of the velocity map, and we use these variations for our uncertainty. Kinematic misalignment larger than 10\degr\,was already observed for galaxies of the SAURON project by \cite{davor10}, but mostly in triaxial, slow-rotating galaxies. In fast-rotating galaxies, kinematic misalignment is only observed in combination with triaxial structures like bars or  shells, or in the very centre of the galaxy.

In a first order approximation, the potential is spherically symmetric because a supermassive black hole is embedded in the nuclear star cluster. In a spherically symmetric system  stellar orbits are approximately planar rosettes at all orientations. But  the sense of rotation of some stars on inclined orbits could be reversed. Then one would observe an asymmetric velocity field.  The same is valid for triaxial or flattened potentials. Triaxial  tumbling systems that extend beyond the radius of influence of a central black hole have different sequences of orbits \citep{heisler82}. If those orbits are populated equally, the system  can be in dynamical equilibrium \citep{schwarzschild82}. Then the velocity field  appears symmetric relative to the short axis when observed from the Sun's position inside the Galactic disk. However, asymmetries can be observed if a subset of inclined stellar orbits is preferentially populated.

Further, there appears to be a substructure that is red-shifted to the Galactic North ($v\,\!\sim\,\!35$\,km/s), and blue-shifted to the Galactic South ($v\,\!\sim\,\!-25$\,km/s). This substructure rotates approximately perpendicular to the large scale sense of rotation. 
The sense of rotation of the well defined clockwise disk  \citep[e.g.][]{levin03,lu09,bartko09,yelda14} is approximately opposite to the rotation we see in our data.  But as we observe rather old stars, which show CO absorption lines, and the clockwise disk is seen in young stars, we do not expect the same dynamical properties.

\cite{tremaine75} suggested that nuclear star clusters form from infalling globular clusters. The kinematic misalignment and the perpendicular substructure  could be the debris of such  accretion events. \cite{antonini12} give a formula for the disruption radius $r_{disr}$ at which an infalling massive star cluster is disrupted due to tidal stresses from the supermassive black hole (SMBH) in the nuclear star cluster. They found $r_{disr} \approx $ 1\,pc for the Milky Way nuclear star cluster, which is roughly equal to the core radius $r_c$ of the old stellar distribution \citep[0.5\,pc,][]{popyoungphot}. 
 \cite{antonini13} used an analytical model for globular cluster infall and found that the initial mass of the infalling clusters has to be at least 10$^7$\,M$_\odot$ to penetrate to a galactocentric radius of $\sim$2\,pc. However, this model neglects internal dynamics of the infalling cluster like e.g. mass segregation.  
Taking this into accout  allows the infalling cluster to bring stars even closer to the centre  \citep{antonini14}. Therefore a substructure at $\sim$0.8\,pc, like the one our data indicates,  could be the remnant of a tidal disruption event. 
 
 \cite{antonini14} ran $N$-body simulations to investigate the consecutive inspirals of  12 clusters to a Galactic centre. In this simulation only the orbits of stars from the first seven infalling clusters are close to isotropic at the end of the simulations. This is not the case for the orbits of stars from the last infalling clusters. After $\sim$3\,Gyr the stellar orbits  are still largely correlated, and would  require about 10\,Gyr to reach a fully isotropic distribution  (F. Antonini, personal communication, April 2014). The relaxation time of the nuclear star cluster is several Gyr at all radii, and outside of 1\,pc, the relaxation time is even longer than the age of the universe \citep{merrittrelax}. Therefore we  expect the signatures of infalling clusters to remain coherent over a long period of time.
 
Our observation of a kinematic misalignment and a perpendicular rotating substructure can be explained by the accretion of massive star clusters. 
 This would  support the theory that  the accretion of massive stellar cluster plays a role in the formation of nuclear star clusters. We plan follow-up observation with KMOS (VLT) to verify  these discoveries.

\subsection{Underestimation of the black hole mass}
 Our two-dimensional Jeans models in Section~\ref{sec:sbjeans} resulted in a black hole mass that is lower by a factor $\sim$2  than  the result from direct measurements of $\sim$\:\!4~$\times$~10$^6$~M${_\odot}$ \citep{ghez08,gillessen09}. Before those measurements, the black hole mass was underestimated by other studies as well \citep{mcginn89,krabbe_sfh,haller96,genzel96}.

As shown in Figure~\ref{fig:dprof}, the ISAAC data velocity dispersion   values are lower than the proper motion data of \cite{Rainerpm09}. This could be due to a  bias in the measurement  of the velocity dispersion in the ISAAC data, and cause the underestimation of the black hole mass.  We  derive the  kinematics  from  the radial velocities of  old stars only, and  cannot trace the young population of stars in the centre. Therefore the measured velocity dispersion is mostly from stars that lie at a large distance in front and behind of Sgr~A*, and only appear to be close to the SMBH in projection \citep{sellgren90}. This means that the measured velocity dispersion of the central $\sim$8\arcsec\,(0.3\,pc) is rather too low \citep[e.g.][]{haller96}.

But as the radius of influence of the black hole  extends out to 2.3\,pc ($\sim$60\arcsec, see Section~\ref{sec:massprofile}) and the young stellar population extends only to 0.5\,pc, this may
 not the only reason for the too low black hole mass. To test this assumption we excluded the kinematic data of the centre out to 0.8\,pc (20\arcsec) and found no changes in the best-fit parameters of the Jeans models.  We also considered the  extreme and unphysical  case that  the value of M/L is zero for the Gaussian components of the potential with $\sigma_{MGE} \le$ 0.8\,pc. This is motivated by the fact that the M/L of  young stars, which are in the central 0.5\,pc, is lower than the M/L of old stars. This increased the black hole mass, but only to 3.4\,\!$\times$\,\!10$^6$\,M$_\odot$. 

Therefore the stellar population change  alone  cannot explain the too low black hole mass.
The position angle offset may also have an effect on the black hole mass measurement in the two-dimensional Jeans models. Further, there might be a systematic bias in black hole mass measurements obtained with integrated light measurements, leading to an underestimation of black hole masses also in  some other galaxies.  We plan to investigate this issue in more detail in a follow-up paper.

\subsection{Mass profile of the cluster}
\label{sec:massprofile}
We obtain the cluster mass from the surface brightness assuming  a constant M/L. As the stellar population in the nuclear star cluster is not uniform, this may be an oversimplification. But population synthesis models showed that, compared  to optical light, the mid-infrared mass-to-light ratio is rather constant.  Under this assumption, \cite{ml} found a mean value of (0.5$\pm$0.1) $M_\odot/L_{\odot,3.6 \mu m}$ in disk galaxies.  \cite{meidt} constrained the   mean value to 0.6\,$M_\odot/L_{\odot,3.6 \mu m}$. The mass-to-light ratio at 4.5\,$\mu m$ should be similar or even less than the mass-to-light ratio at 3.6\,$\mu m$  \citep{oh}. This is in agreement with our results from the Jeans models of  0.56$^{+0.22}_{-0.26}$\,$M_\odot/L_{\odot,4.5\,\mu m}$. However, our best fit black hole mass is too low compared to direct measurements, and this might influence the outcome of the M/L. We tested the magnitude of this by using the black hole mass of 4\,$\times$~10$^6$~M${_\odot}$ as input. Then the M/L decreases to 0.35, which is still above our lower limit value.  
This test shows that our result of M/L is robust under local variations of the stellar populations, since the kinematic data  covers a large enough region.

\begin{figure}
\resizebox{\hsize}{!}{\includegraphics{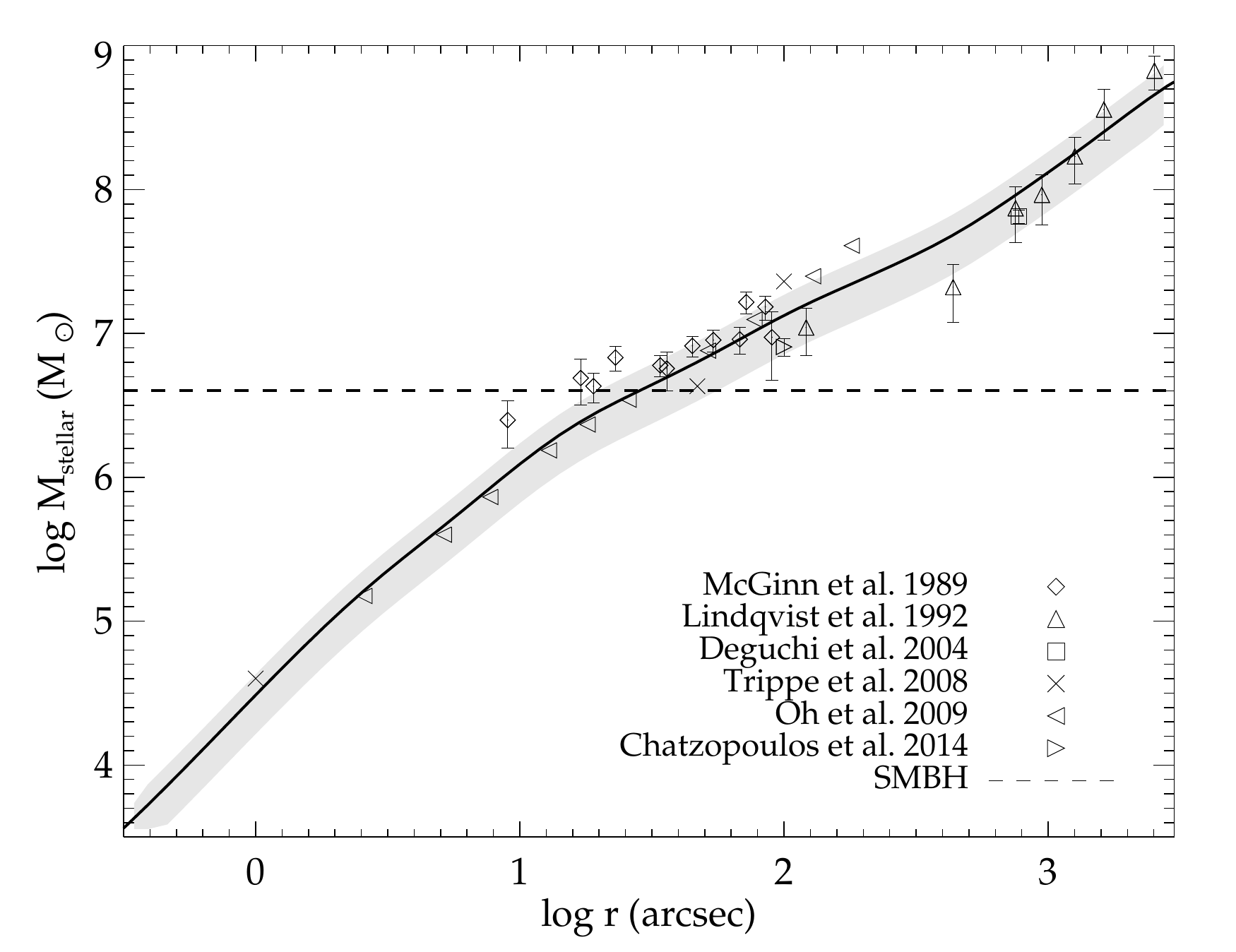}}
\caption{Enclosed stellar mass  within a distance of 0.3\arcsec\,to 50\arcmin\,along the mean radius of the ellipses in units of $M_{\odot}$ and in logarithmic scaling. The  black line denotes a M/L value of 0.56, the grey shaded contours show the uncertainty in M/L of $^{+0.22}_{-0.26}$. The horizontal line is for a supermassive black hole with the mass M$_\bullet$ = 4~$\times$~10$^6$~M${_\odot}$. We also plot the  results for the enclosed mass from previous studies: 
\citealt{mcginn89} (assumed Galactocentric distance $R_0$ = 8.5\,kpc), \citealt{lindqvist922} ($R_0$ = 8.5\,kpc), \citealt{deguchi} ($R_0$ = 8\,kpc), \citealt{trippe08} ($R_0$ = 8\,kpc), \citealt{Oh09} ($R_0$ = 8\,kpc) and \citealt{chatzopoulos} ($R_0$ = 8.3\,kpc).
}
\label{fig:encmass}
\end{figure}

Our enclosed mass profile is shown in Figure~\ref{fig:encmass} using our best fit M/L of  0.56$^{+0.22}_{-0.26}$\,$M_\odot/L_{\odot,4.5\,\mu m}$ for 0.3\arcsec\textless\,r\,\textless\,50\arcmin. This plot contains the enclosed mass from both the nuclear star cluster and the nuclear stellar disk, which dominates the mass at r\,$\gtrsim$\,100\,pc (40\arcmin).  
We also plot the results for the enclosed mass for various other studies, among those  \cite{mcginn89} and  \cite{lindqvist922}  in Figure~\ref{fig:encmass}. Those two studies  assumed a Galactocentric distance of 8.5\,kpc, while we assume 8.0\,kpc. Further, they compute the  mass by integrating the light enclosed in  circles, and not ellipses, and assume spherical symmetry. These are probably more important effects. 
The results of \cite{mcginn89}  for the enclosed mass tend to be higher at r\,\textless\,30\arcsec, but we have good agreement with \cite{mcginn89} at larger radii, and also with the results of \cite{lindqvist922}. 
One out of three of the values given by  \cite{trippe08} is higher than our upper limit, while the other two data points agree with our results. 
 The results of \cite{Oh09} are in excellent agreement with our data inside 100\arcsec. 
At 28$^{+28}_{-9}$\arcsec\,(1.1$^{+1.1}_{-0.3}$\,pc), the mass of the stars equals the mass of the SMBH. Inside this radius the potential  is dominated by the SMBH and therefore close to spherically symmetric. \cite{merritt04} defines the radius of influence $r_{infl}$ of a SMBH as the radius at which the enclosed mass in stars is twice the mass of the black hole. With this definition we obtain a mean radius $r_{infl}$ =  60$^{+54}_{-17}$\arcsec\,(2.3$^{+2.1}_{-0.7}$\,pc). This value is in agreement with the result of $\sim$~3\,pc found by \cite{alexander05}, based on the enclosed mass of \cite{rainer03}.

\begin{figure}
\resizebox{\hsize}{!}{\includegraphics{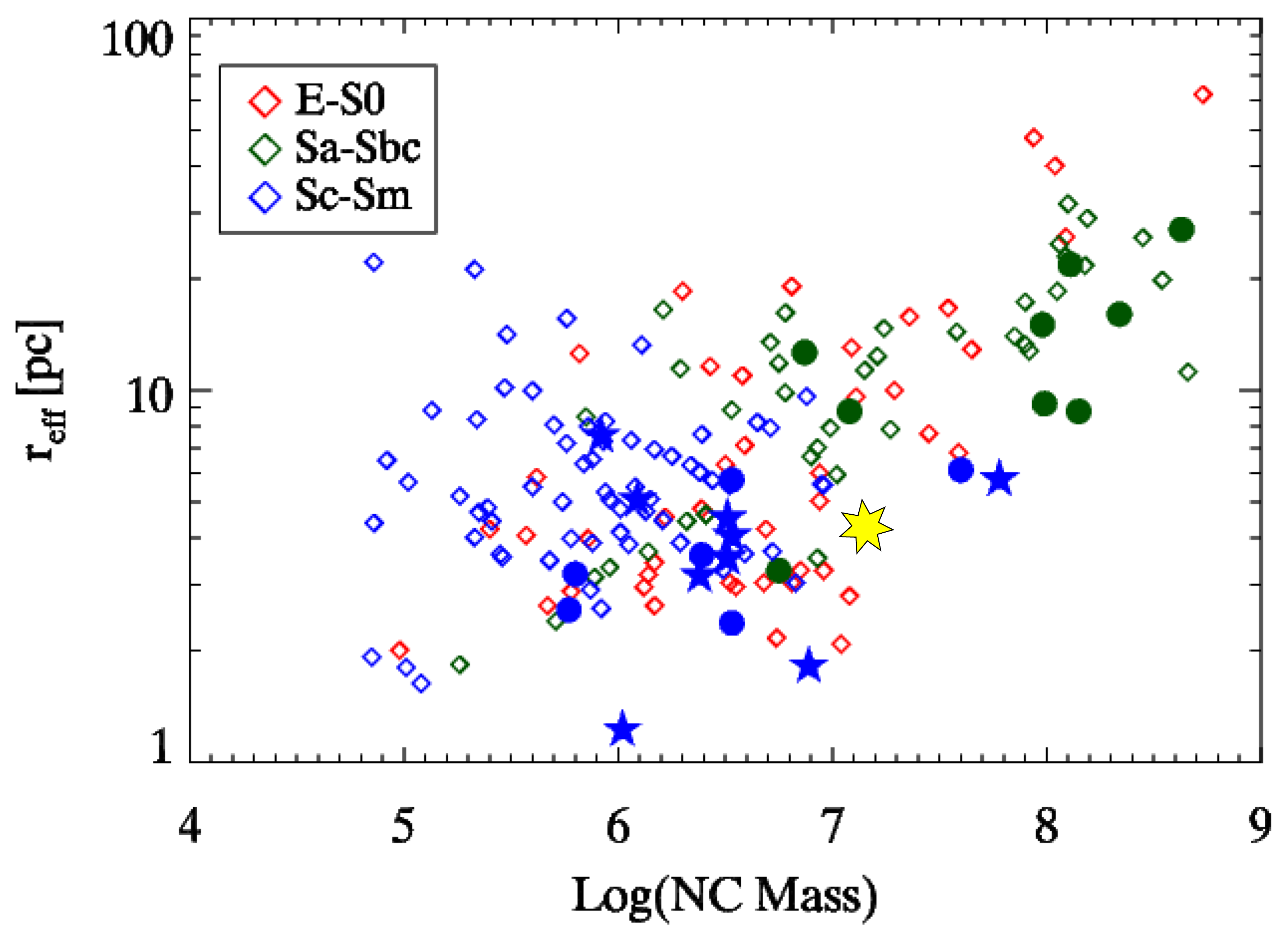} }
\caption{Nuclear cluster mass $M_{NSC}$ $-$ effective radius $r_{eff}$ relation for nuclear star clusters, based on \cite{anil08} with data from  \citet{jakob05,rossa06} shown as filled symbols and \citet{boker02,carollo97,carollo98,carollo02,cote06} as open symbols. The yellow star is the result for the Milky Way nuclear star cluster  within an effective radius of 4.2\,pc \citep{sb}.  }
\label{fig:reffmnsc}
\end{figure}

Studies of \cite{cote06} and \cite{forbes08} showed that for nuclear star clusters  the effective radii increase with the luminosity of the nuclear star cluster. This indicates also a correlation of the nuclear star cluster mass $M_{NSC}$ to $r_{eff}$. 
We plot this relation for a large number of nuclear star clusters compiled by \cite{anil08} in Figure \ref{fig:reffmnsc}; solid points represent the most reliable mass estimates from \cite{jakob05} and \cite{rossa06}.  Other data points are from \cite{boker02,carollo97,carollo98,carollo02,cote06} and the mass estimate was derived purely photometrically.
The yellow star denotes the  mass of the Milky Way nuclear star cluster of 1.4~$\times$~10$^7$\,M$_\odot$ within the effective radius. The $M_{NSC}$ $-$  $r_{eff}$  relation breaks down for low masses. The Milky Way nuclear star cluster is within the trends defined by other clusters, but is somewhat more compact than average.

\subsection{Anisotropy}

\cite{Rainerpm09} found that the velocity dispersion of the late-type stars in the cluster is consistent with  isotropy. This was confirmed by \cite{tuan} for the inner 0.5\,pc of the Galaxy.  
We assumed different radial anisotropy profiles, but there is no clear trend about which profile fits best. However, the models  favour tangential anisotropy, and if we impose constant anisotropy throughout the model, the best-fit value is   $-0.3^{+0.3}_{-0.4}$, i.e. radial anisotropy is excluded.
Also our analysis of $\lambda_R$ suggests  the Milky Way nuclear star cluster is anisotropic.  Specifically, the cluster appears to be more flattened than expected based on the observed rotation.

\subsection{Extreme velocities for individual stars}
We extracted 1,375 spectra from bright stars with K$_S$~\textless~11.5$^m$ from our unbinned data and fitted their velocities in Section~\ref{sec:co}. 
For three stars we obtained velocities higher than 250\,km/s, they are listed in Table~\ref{tab:sphe}. One is already known as IRS~9, and at a projected distance of 8\arcsec\,from Sgr~A* with $v = -340$\,km/s. Its velocity was already measured by \cite{rv_fast_08} to $-347.8$\,km/s. The other stars are further out, at 80\arcsec\,and 130\arcsec\,distance, and have radial velocities of 292\,km/s and $-266$\,km/s. This corresponds to  3.5$\sigma$ and 3.8$\sigma$  in the velocity distribution of individual stars.  These extreme velocities are even more surprising in their position as the local velocity dispersion around these stars is low. It is possible that these stars  have been accelerated by the Hills mechanism. 
\cite{hills} predicted the existence of hypervelocity stars, which were ejected in a three-body encounter between the SMBH and a binary star system. Depending on the semi-major axis distance of the binary and the closest approach to the SMBH, stars can be accelerated to velocities of some 1,000\,km/s. Since the  stars detected by us have velocities of $-266$\,km/s and 292\,km/s, they would require a proper motion component of $\sim$960\,km/s to reach a magnitude of the  velocity vector of $\sim$1,000\,km/s.  Therefore we do not consider them as strong hypervelocity star candidates. Nevertheless, these stars could have been accelerated by the Hills mechanism.

\section{Conclusions}
\label{sec:conclusions}
We obtained an integral field likeÓ spectroscopic data set of the Milky Way nuclear star cluster. This data covers the central 11\,$arcmin^2$ ($\sim$~60\,pc$^2$), and six smaller  fields at larger projected distances. Those go  out to 8\arcmin\,distance from Sgr~A* along the Galactic Plane and cover 3.2\,$arcmin^2$ in total. We set up  a spectroscopic map with a pixel size of 2\farcs22 of the central field. 
Additionally we constructed a spectroscopic map that was cleaned of bright stars and foreground stars.  Using these data cubes, we fitted the stellar CO absorption lines and computed velocity and velocity dispersion maps. 
\begin{enumerate}

\item 

We found a misalignment of the photometric and kinematic position angles by $\sim$\,\!9\degr\,Galactic east of north. 
Further, we  detected indications for a new kinematic  substructure that is approximately aligned along the Galactic minor axis at $\sim$\,\!20\arcsec\,(0.8\,pc) distance from Sgr~A*.  Both observations indicate to  different accretion events of the Milky Way nuclear star cluster. To confirm these findings with higher spatial resolution  and a wider wavelength range  we plan new observations with  KMOS (VLT).  If confirmed, the position angle offset  and the substructure give important clues to the formation history of nuclear star clusters. They support the theory that infalling massive clusters  play a role in the build-up of nuclear star clusters.

\item
Axisymmetric Jeans models of the velocity maps underestimated the mass of the supermassive black hole in the Galactic centre. The reason for this can partially be explained by the lack of old stars in the central 0.5\,pc. Therefore our measured velocity dispersion was biased to too low values. Another possibility is that the kinematic misalignment influences the outcome of the models. We cannot exclude  a systematic bias in black hole mass measurements obtained using integrated light. Such a bias could mean that we underestimate black hole masses in some other galaxies as well. 
Therefore we plan to investigate this issue further using additional data in a follow-up paper. 

\item 
We fitted a surface brightness profile of the Milky Way nuclear star cluster using  NACO $K$-band data and \emph{Spitzer} photometry at 4.5\,$\mu m$. Our best-fit Jeans models resulted in   a mass-to-light ratio  M/L$_{4.5\,\mu m} $ = 0.56$^{+0.22}_{-0.26}$\,$M_\odot/L_{\odot,4.5\,\mu m}$. From these results we computed a profile for the enclosed mass. 
At a distance of $r_{eff}$ = 4.2\,pc from Sgr~A* the cluster mass (without the black hole mass) is (1.4$^{+0.6}_{-0.7}$)\,$\times$~10$^7$~M${_\odot}$. Compared to nuclear star clusters of similar size the Milky Way nuclear star cluster is rather massive.

\end{enumerate}
 \begin{acknowledgements}
 This research was supported by the DFG cluster of excellence Origin and Structure of the Universe (www.universe-cluster.de).
 This publication makes use of data products from the Two Micron All Sky Survey, which is a joint project of the University of Massachusetts and the Infrared Processing and Analysis Center/California Institute of Technology, funded by the National Aeronautics and Space Administration and the National Science Foundation. This research made use of the SIMBAD database (operated at CDS, Strasbourg, France). We would  like to thank the ESO staff who helped us to prepare our observations and obtain the data. A special thanks to our Telescope operator J. Navarrete, who implemented all our non-standard  observing  techniques. A. F. also thanks Holger Baumgardt and Eric Emsellem for
  helpful discussions about the project. We thank Fabio Antonini for further investigations of his simulations on our behalf and providing us additional information.  C. J. W. acknowledges support through the Marie Curie Career Integration Grant 303912. We finally thank the anonymous referee for  useful comments and suggestions.
 \end{acknowledgements}
 \bibliographystyle{aa}		

\bibliography{bibs}		
\newpage
\begin{appendix}
\section{Persistence removal} 
\label{sec:pers}
Persistence is a known problem of infrared detectors when observing bright sources. If the exposure time  is too long and the source is overexposed, there can be ghost images of the source in a subsequently taken exposure of a faint source. The magnitude of the  persistence effect decreases in time according to a power-law \citep{acsbook}. To avoid persistence, one should either choose short exposure times or, alternatively, make a sequence of read-outs afterwards, to flush the detector's memory of the bright source.  
We took acquisition images right before the object spectra, which lead to persistence in the 2D spectra. Also our 2D sky spectra are affected as we took three dithered images on source before the sky offsets to verify the position on the sky after the drift, and two dithered images on sky {that also contain a few stars}. 

However, we developed a way to remove the persistence signature completely in our sky frames and to some extent also in the object spectra.     We use the images that cause the persistence and the corresponding 2D spectra that are affected by the persistence.  Removing  persistence from the 2D sky spectra  is rather straight forward: For every 2D sky spectrum $S_{raw,sky}$ we compute the median of each column along the dispersion axis    and subtract it from each pixel in the column, and  do the same with the  median along the spatial axis to get a 2D  sky spectrum without any lines or stars $S_{med,sky}$. The only features left in the median subtracted  2D  sky spectrum are from the persistence.  Then one can simply compute a corrected 2D  sky spectrum $S_{cor,sky}$
\begin{equation}
S_{cor,sky}  =  S_{raw,sky} - S_{med,sky} {+} mean(S_{med,sky}).
\end{equation}
 The clean 2D  sky spectra are combined to 2D  Mastersky spectra using IRAF.  We also produce noise files for each sky file that contain the information about the persistence correction.

This simple approach is not possible for the 2D  object spectra, as there are many stars with strong absorption lines and also  an H$_2$ gas emission line. In contrast to the 2D  sky spectra, subtracting a median leaves other strong features apart from the  persistence residuals. Therefore we decided to model the persistence and subtract the model from the 2D  object spectra. To determine the persistence model parameters we use the 2D sky spectra before we apply the aforementioned correction on them. 

Saturated pixels have negative counts in our images and leave persistence features in the 2D spectra taken afterwards. Persistence is also a problem if  the counts of a bright source  at a given pixel of the detector are above a certain threshold. First, one needs to determine the value of this threshold. Therefore we consider only counts above a trial threshold and make a persistence mask $M$, where pixel with negative counts in the image as well as counts above the threshold are set to one and all other pixels are set to zero. As the  raw 2D sky frames are affected by persistence coming from three images on source and two images  on sky, taken $\sim$\:\!2$-$3 minutes later, we make two masks, one for the images on source ($M_{source}$) and one for the sky images ($M_{sky}$). Then we use  \emph{mpfit2dfun.pro} \citep{mpfit} to fit the amplitudes  of the persistence, $K_{source}$  and $K_{sky}$,   and a full-width-half-maximum FWHM for a Gaussian smoothing filter $G_{FWHM}$ of the masks. The residual spectrum $R$ is  
\begin{equation}
R = S_{med} - G_{FWHM} \ast (K_{source} \times M_{source} + K_{sky} \times M_{sky} ),
\end{equation}
where the symbol ``$\ast$'' denotes convolution.
 We  try  different values of the mask threshold, and we find a minimum of the standard deviation of the residual spectrum $R$ at a threshold of 33,500 counts. Therefore we define 33,500 counts as our threshold for further corrections of images taken with an exposure time of 20$s$. For the images with exposure time $t= 5s$, we have a lower threshold of 24,000 counts, found by the same method. 

The next step is to  fit $K_{source}$,  $K_{sky}$, and FWHM with the chosen threshold value of the mask for every median subtracted  2D sky spectrum $S_{med,sky}$. The persistence signal decreases in time, therefore  $K_{source}$ and   $K_{sky}$ decrease from the first to the fifth 2D sky spectrum in a sequence. We fit the decrease of  $K$  with the power law  
\begin{equation}
K(t) = A \times t^{-\gamma},
\end{equation} 
where the parameter $A$ is  the amplitude and   $\gamma$ is a scale factor. We do this for  $K_{source}$ and   $K_{sky}$ seperately and together, but our later 
analysis shows that we obtain better results when  we use the result of fitting  $K_{source}(t)$ alone. The  parameters we use are  $A = (6900\,\!\pm\,\!1700)$ and $\gamma = (0.98\,\!\pm\,\!0.04)$. The uncertainties are the formal 1-$\sigma$ errors of the fit.  With this knowledge we can subtract the persistence  from the 2D object spectra with the equation 
\begin{equation}
S_{cor,object} = S_{raw,object} -  (G_{FWHM} \ast M_{acqu})   \times A  t^{-\gamma}.
\end{equation}
The images causing persistence in the 2D  object spectra are acquisition images.  As we made acquisition offsets, we subtract  persistence   caused by the acquisition image itself  and from a shifted acquisition image. We have only the acquisition  image itself as file, but we can identify the shift in pixels from the persistence in the 2D  spectra, and we can fit the shift of the acquisition offset together with  an attenuation factor $\alpha$. This factor $\alpha$ is necessary as the offset was performed before the acquisition image was taken, and therefore the persistence signal is weaker. In this case the corrected 2D  object spectra $S_{cor,object}$ are  
 \begin{equation}
\begin{split}
 S_{cor,object} = S_{raw,object} -  (G_{FWHM} \ast M_{acqu})   \times A  t^{-\gamma}  \\ 
 		-  \alpha \times (G_{FWHM} \ast M_{acqu,shifted})  \times A  t^{-\gamma}.
 \end{split}
 \end{equation}
 
 With  these procedures it is possible to remove the persistence structures completely from the 2D   sky frames and to subtract them from  the 2D   object spectra partially, but some residuals still remain.  In some cases  we made more than one acquisition offset, and furthermore   it is difficult to model the shape of the persistence, which was slightly elliptical due to the drift during the acquisition. The shape also  changes depending on the area on the detector  and seeing conditions. To account for uncertainties in the persistence subtraction, we make noise files that are used for the further analysis of the spectra. 
 
To check the influence of the persistence after the correction on our results we compare two  2D spectra that cover almost the same region of the Milky Way nuclear star cluster. One 2D spectrum was not affected by the persistence, since it was the   20th exposures taken after the acquisition image. The other 2D spectrum was taken shortly after the acquisition image and was affected by persistence. We extract several 1D spectra by summing between 45 and 100 rows of the 2D spectra. Then we fit the CO absorption lines with \emph{ppxf}. The signal-to-noise ratio (S/N) of the spectra taken from the file with persistence, but after our correction, is lower by $\sim$27\% compared to the S/N of spectra from the file with no persistence. But as the effect of persistence decreases in time, not all of our data are as much affected. We estimate that in $\sim$70\% of our spectra the decrease in S/N due to the persistence is less than 20\%.

\section{H$_2$ gas emission Kinematics}
\label{sec:h2}

\begin{figure}
\resizebox{\hsize}{!}{\includegraphics{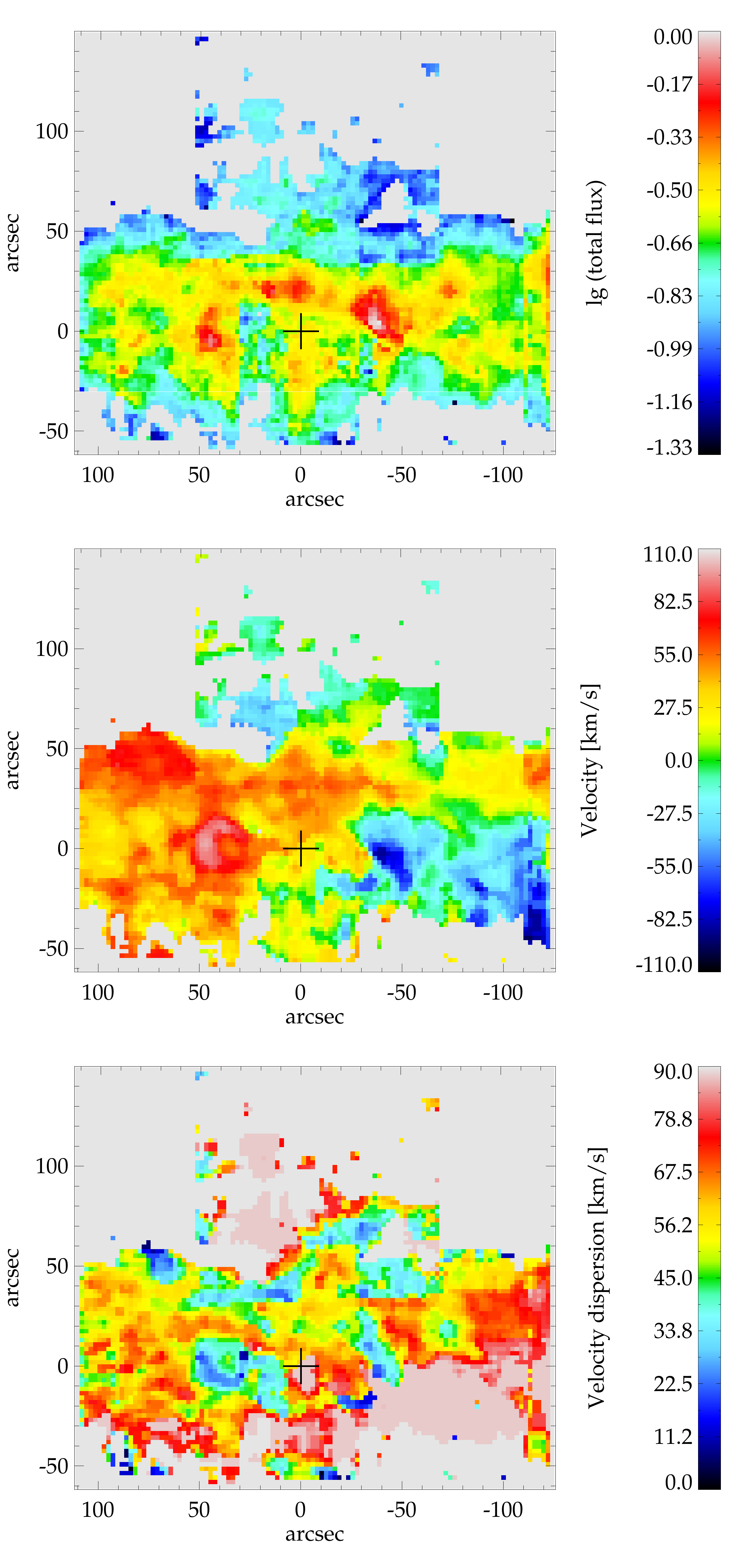} }
\caption{Results of the single Gaussian fit to  the H$_2$ gas emission line. Upper panel: flux  in logarithmic scaling and with respect to the maximum flux. Middle panel: velocity in km/s. Lower panel: velocity dispersion in km/s,  corrected for the instrumental dispersion ($\sigma_{instr} \approx$ 27\,km/s).  The coordinates are centred on Sgr~A* and along the Galactic plane with a position angle  of 31\fdg40. The cross marks the position of Sgr  A*.  }
\label{fig:h2all}
\end{figure}

We fitted the 1-0 Q(1) 2.4066~$\mu m$ transition of H$_2$ to make a comparison with previous studies \cite[e.g.][]{gatley86,yusefh2,Lee08}.  
The excellent agreement of our flux and kinematic maps with these studies shows that our data can reproduce previous results without strong biases.
Figure~\ref{fig:h2all} illustrates the results of the fitting for flux, velocity and velocity dispersion. Regions where the flux was too low and  confusion with sky line residuals might be possible  are not displayed. 
In the H$_2$ velocity map  we can identify the northeastern and southwestern lobe of the circumnuclear disk (see Figure~1 of \cite{amobaladron11} for an illustration of the central 12 parsec). 

\end{appendix}

\end{document}